\documentclass[reprint,amsmath,amssymb,aps,prb]{revtex4-2}

\usepackage{graphicx}
\graphicspath{{figures/}}
\usepackage{dcolumn}
\usepackage{bm}
\usepackage{ragged2e}
\usepackage{hyperref}
\hypersetup{
           breaklinks=true,   
           colorlinks=true,   
           allcolors =RoyalPurple 
        }
\usepackage{dsfont}
\usepackage{mathrsfs}
\usepackage{wrapfig}
\usepackage{physics}
\usepackage{subcaption}
\usepackage{cancel}
\usepackage{verbatim}
\usepackage{dsfont}

\usepackage[usenames,dvipsnames]{xcolor}
\usepackage{amsmath}

\begin{document}

\title{Slave-spin approach to the Anderson-Josephson quantum dot}
\author{Andriani Keliri}\email{andriani.keliri@college-de-france.fr}
\affiliation{JEIP, UAR 3573 CNRS, Coll\`{e}ge de France, PSL Research University,
11, place Marcelin Berthelot,75231 Paris Cedex 05, France}
\author{Marco Schir\`{o}}
\affiliation{JEIP, UAR 3573 CNRS, Coll\`{e}ge de France, PSL Research University,
11, place Marcelin Berthelot,75231 Paris Cedex 05, France}

\date{\today}

\begin{abstract}
We study a strongly interacting quantum dot connected to two superconducting leads using a slave-spin representation of the dot. At the mean-field level, the problem maps to a resonant level model with superconducting leads, coupled to an auxiliary spin-1/2 variable accounting for the parity of the dot. We obtain the mean-field phase diagram, showing a transition between a Kondo (singlet) and a local moment (doublet) regime, corresponding to the $0-\pi$ transition of the junction. The mean-field theory qualitatively captures the Kondo singlet phase and its competition with superconductivity for weak values of the BCS gap, including the non-trivial dependence of the Andreev bound states on the interaction, but fails in the doublet regime where it predicts a dot decoupled from the bath. Using diagrammatic techniques and a random phase approximation, we include fluctuations on top of the mean-field theory to describe finite-frequency dynamics of the effective spin variable. This leads to the formation of high-energy Hubbard bands in the spectral function and a coherent Kondo peak with a BCS gap at low energies. We compute the Josephson current and the induced superconducting correlations on the dot. Finally, we evaluate the microwave response in the strongly interacting Kondo regime.
\end{abstract}

\maketitle

\section{Introduction}
The Kondo effect and superconductivity are two paradigmatic examples of many-body phenomena in which the electron spin plays an important role. In the Kondo effect \cite{kondo_resistance_1964}, the spin of a magnetic impurity immersed in a Fermi sea can be screened by the conduction electrons, resulting in the formation of a spin singlet. As a result, the impurity behaves as nonmagnetic below the Kondo temperature $T_K.$ In the Bardeen-Cooper-Schrieffer (BCS) theory of superconductivity \cite{BCS}, electrons form pairs with opposite spins, resulting in a singlet state whose binding energy per electron is equal to the superconducting gap $\Delta$ \cite{Cooper}. An experimentally accessible way to bring about the interplay of these two phenomena is to couple a quantum dot (QD) to superconducting leads (S) \cite{hybridsystems, buitelaar_quantum_2002,cleuziou_carbon_2006,jarillo-herrero_quantum_2006,vandam_supercurrent_2006,buizert_kondo_2007,jorgensen_critical,eichler_tuning_2009,deacon_tunneling_2010,lee_zerobias_2012,maurand_first-order,pillet_tunneling_2013,delagrange_manipulating_2015,delagrange_2016,jellinggaard_tuning_2016, lee_scaling_2017}.

A QD is a small electronic system that acts as an artificial atom \cite{dots,dots2}. Because of the spatial confinement of electrons in the QD, its resulting energy levels are discrete. Provided that the level spacing is sufficiently large, the QD can be approximated by a single discrete level whose position can be tuned with a gate voltage. In such a small system, the effect of the Coulomb repulsion $U$ can become important, modifying electron transport through the dot (leading, for example, to Coulomb blockade). When the single level is occupied by only one electron, the QD resembles a magnetic impurity and is theoretically described by the single-level Anderson impurity model (AIM) \cite{anderson_localized_1961}. When connecting the QD to some metallic leads, one expects the Kondo effect to affect its transport properties \cite{cronenwett_tunable_1998,goldhaber-gordon_kondo_1998,kouwenhoven_revival_2001,pustilnik_kondo_2004}. Indeed, the Kondo effect leads to the formation of a resonance in the density of states of the QD, pinned at the Fermi energy and having a width of $k_BT_K.$  The Kondo resonance facilitates transport through the dot, making it completely transparent when the temperature is lowered below $T_K.$

When the leads are superconducting, a gap opens in the density of states of the QD, and a pair of bound states, the Andreev bound states (ABS), form inside the gap. Provided there is a finite phase difference between the leads, the Josephson effect leads to a supercurrent flowing through the QD, mostly carried by the ABS \cite{martin-rodero_josephson_2011,controllable_2007}. The ground state of the system depends on the competition between the Kondo correlations, favoring the formation of a Kondo singlet, the superconducting correlations, favoring the formation of a BCS singlet, and the Coulomb interaction, favoring a magnetic doublet state. A singlet-doublet transition becomes possible by tuning the various parameters of the system (the coupling strength to the leads, the energy of the QD level, or the superconducting phase difference between the leads). In the doublet phase, the Josephson current abruptly changes sign and is suppressed \cite{vandam_supercurrent_2006,cleuziou_carbon_2006}, so that the S-QD-S system behaves as a so-called ``$\pi-$junction", in contrast to the non-interacting ``$0-$junction" behavior. The $0-\pi$ phase transition is accompanied by the crossing of the ABS at zero energy, expressing the fact that the ABS correspond to transitions between the ground state and the excited state \cite{meng_selfconsistent_2009}.

Recently, renewed interest has emerged around the physics of the superconducting AIM. This is largely due to the idea of using the S-QD-S system as a platform to realize Andreev-level and Andreev spin qubits \cite{PhysRevB.64.140511,andreevqubit,coherent,PhysRevLett.90.226806,PhysRevB.81.144519,doi:10.1126/science.abf0345}, as well as qubits using the doublet ground state in QD Josephson junctions \cite{bargerbos2022singlet,pita-vidal_direct_2023}. Another motivation is the potential realization of topological qubits hosting Majorana bound states \cite{sau2012realizing,PhysRevB.86.134528}. Recent experimental investigations of poor-man Majorana chains, for example, are based on models of quantum dots coupled to superconducting leads \cite{dvir_realization_2023}, which reduce, in a suitable parameter regime, to the Kitaev model \cite{kitaev2001unpaired}. The development of various experimental techniques that have allowed the experimental observation of ABS in QDs using tunneling spectroscopy \cite{pillet_tunneling_2013}, supercurrent and Josephson spectroscopy \cite{PhysRevX.3.041034,griesmar_superconducting,peugeot_twotone_2024}, or microwave spectroscopy \cite{fatemi_microwave_2022,kurilovich_microwave_2021,matute2022signatures, sahu_groundstate_2024}, provides further motivation for a better theoretical understanding~\cite{moca2021kondo,pavesic2024strong,pavesic2023impurity}. An interesting question concerns the dynamics of the S-QD-S system and its response to external drives, such as the non-linear transport under Floquet drive and the physics of multiple Andreev reflections~\cite{keliri2023driven,keliri2023longrange}, recently also observed in a quantum transport emulator based on ultracold atoms~\cite{huang2023superfluid,visuri2023dc}.

In light of these motivations, it is interesting to explore theoretical approaches to the superconducting AIM, that can be then extended to a non-equilibrium regime. Traditionally, the physics of the superconducting AIM in equilibrium has been studied with different analytical approaches~\cite{martin-rodero_josephson_2011,meden_anderson_2019,seoanesouto_subgap_2024}, including perturbative methods~\cite{glazman_resonant_1989,vecino_2003,zonda_perturbation_2016}, mean-field and slave-boson approaches~
\cite{rozhkov_josephson_1999,vecino_2003,borkowski_lowtemperature_1994,schwab_andreev_1999,rozhkov_interactingimpurity_2000,bergeret_josephson_2007}, as well as within a non-crossing approximation~\cite{clerk_2000,sellier2005}, each of them working in specific parameter regimes. On the numerical front, methods such as the numerical renormalization group (NRG) \cite{yoshioka_numerical_2000,choi_kondo_2004,tanaka_kondo_2007,bauer_spectral_2007,karrasch_josephson_2008} and finite-temperature quantum Monte-Carlo \cite{luitz_weakcoupling_2010,luitz_understanding_2012} can provide quantitative comparison to experiments in equilibrium, but are computationally expensive and their extension to the dynamical regime remains an open question. Moreover, these numerical methods are difficult to extend to the study of more complex structures, such as multiterminal and multidot junctions.

In this paper, we use a slave-spin representation to study the superconducting AIM semi-analytically. The class of slave-particle methods was designed to tackle strongly interacting effects at the mean-field level, and can capture the formation of the Kondo resonance. The slave-spin approach has proven useful for the study of the Hubbard model~\cite{demedici_orbitalselective_2005,huber_dynamically_2009,ruegg_2010,schiro_quantum_2011,zitko_2015}, as well as the AIM~\cite{baruselli_subohmic_2012,guerci_unbinding_2017}, both in and out of equilibrium. It provides a minimal and efficient representation, where the Hilbert space enlargement is just an auxiliary two-level system. This method allows us to map the original interacting problem at half-filling to a non-interacting S-QD-S problem coupled to an Ising spin. Moreover, including spin fluctuations around the mean-field solution allows to build up the high-energy Hubbard bands~\cite{raimondi_lower_1993,guerci_transport_2019}. 

This manuscript is organized as follows. In Sec.~\ref{sec:model} we define the model and its slave-spin representation. In Sec.~\ref{sec:meanfield} we discuss the mean-field theory of the problem and present the resulting phase diagram and spectral properties. We highlight the aspects that are qualitatively captured by a simple mean-field decoupling and those that are not, most notably the description of high-energy incoherent excitations (Hubbard bands). In Sec.~\ref{sec:beyond} we include fluctuations corrections to the slave-spin variable, overcoming the main issue of the mean-field theory and  including corrections to various physical quantities coming from incoherent excitations. In Sec.~\ref{sec:microwave} we calculate the microwave response of the superconducting QD. In Sec.~\ref{sec:conclusions} we present our conclusions. Further technical details are presented in Appendices~\ref{sec:largegap}-\ref{app:microwave}.

\section{The model and its slave-spin representation}\label{sec:model}
We consider an S-QD-S junction, formed by a quantum dot coupled to two superconducting leads. The system is described by the superconducting Anderson impurity model with Hamiltonian
\begin{equation}\label{model}
    \mathcal{H}=\mathcal{H}_{\mathrm{dot}}+\mathcal{H}_{\mathrm{BCS}}+\mathcal{H}_{\mathrm{tun}}.
\end{equation}

The dot consists of a single level with energy $\varepsilon_d$ and an on-site Coulomb interaction $U,$ 
\begin{equation}
    \mathcal{H}_{\mathrm{dot}}=\sum_\sigma (\varepsilon_d+\frac{U}{2}) f^{\dagger}_{\sigma}f_{\sigma}+\frac{U}{2}\pqty{\sum_\sigma f^{\dagger}_{\sigma}f_{\sigma}-1}^2,
\end{equation}
where the operators $f^{\dagger}_\sigma$ and $f_\sigma$ respectively create and annihilate an electron with spin $\sigma$ on the dot. In what follows, we will assume the system is always at the particle-hole symmetric point, corresponding to $\varepsilon_d=-\frac{U}{2}$.

The leads are described by the BCS mean-field Hamiltonian,
\begin{equation}
    \mathcal{H}_{\mathrm{BCS}}= \sum_{jk\sigma} \varepsilon_k c^{\dagger}_{jk\sigma}c_{jk\sigma}+\sum_{jk}(\Delta e^{i\phi_j}c^{\dagger}_{jk\uparrow}c^{\dagger}_{j-k\downarrow}+ \mathrm{H.c.}),
\end{equation}
where the index $j=L,R$ denotes the left or right lead. The operators $c^{\dagger}_{jk\sigma},c_{jk\sigma}$ create and annihilate electrons at lead $j$ with momentum $k$ and spin $\sigma.$ The amplitude of the superconducting gap $\Delta$ is taken to be identical for the two leads, while $\phi_j$ is the superconducting phase of lead $j.$ We assume a phase difference $2\phi=\phi_L-\phi_R$ and, without loss of generality, we choose $\phi_L=-\phi_R=\phi.$

Finally, the last term of the model Hamiltonian (\ref{model}) describes the tunneling between the dot and the leads,
\begin{equation}
   \mathcal{H}_{\mathrm{tun}}=\sum_{jk\sigma} V_j \pqty{c^{\dagger}_{jk\sigma}f_{\sigma}+f^{\dagger}_{\sigma}c_{jk\sigma}}.
\end{equation}
The characteristic energy scales of the problem are the superconducting gap $\Delta,$ the Coulomb interaction $U,$ and the hybridization between the leads and the QD, which we define as $\Gamma=\pi\rho_0 V_L^2+\pi\rho_0 V_R^2,$ where $\rho_0$ is a flat density of states of the leads in the absence of superconductivity. Experimentally, these three quantities are often of the same order of magnitude, making an analytical treatment challenging. 

We will rewrite the Hamiltonian in Eq.~(\ref{model}) at the particle-hole symmetric point by means of a slave-spin representation of the dot operators.  The main idea of any slave-particle method is to enlarge the Hilbert space by introducing auxiliary degrees of freedom, supplemented by a constraint projecting the enlarged Hilbert space back onto the physical one. The advantage of this formulation is that it naturally accounts for the fractionalization of quantum numbers that often occurs in strongly correlated electron systems. In the slave-spin method one enlarges the dot Hilbert space, spanned by the states $\vert n\rangle=\left\{\vert0\rangle,\vert\uparrow
\rangle,\vert\downarrow\rangle,\vert\uparrow\downarrow\right\}$, by adding a spin $1/2$ Ising variable $\sigma^z$ with eigenstates $\sigma^z\vert\pm\rangle=\pm \vert\pm\rangle$, with the value $+$ or $-$ associated respectively to the states with even or odd parity. The enlarged Hilbert space reads $\vert n^d_{\sigma}\rangle\otimes \vert\sigma^z\rangle$ with $\vert n^d_{\sigma}\rangle$ the states associated to an auxiliary spinful fermionic variable. We then introduce a projector onto the physical Hilbert space that reads
\begin{align}
P=\frac{1+\sigma^z\Omega}{2},
\end{align}
with $\Omega=1-2\left(n-1\right)^2$ taking the value $\Omega=\pm 1$ when $n^d=1$ or $n^d=0,2$, respectively. From the above we conclude that in the physical Hilbert space the dot destruction operator is replaced by 
\begin{equation}
   f_{\sigma}=\sigma^x d_\sigma.
\end{equation}
Furthermore, one can show that in the physical Hilbert space, where $P=1$, we have $\sigma^z=\Omega$.
The Hamiltonian (\ref{model}) is then mapped to
\begin{equation}\label{Htotal}
  \mathcal{H}'=\mathcal{H}_{\mathrm{BCS}} +\sigma^x \sum_{jk\sigma} V_j \pqty{c^{\dagger}_{jk\sigma}d_{\sigma}\!+\!\mathrm{H.c.}} +\frac{U}{4}\left(1\!-\!\sigma^z\right).
\end{equation}
While the equivalence between the original AIM and the slave-spin formulation is exact in the presence of the projector $P$, it was shown in Ref.~\cite{schiro_quantum_2011} that at the particle-hole symmetric point the slave-spin constraint is not needed, which largely simplifies the theoretical description.  Finally, we note that it is possible to move away from the particle-hole symmetric point, although we will not consider such a case in the present paper. We note only that in the absence of particle-hole symmetry, one would obtain a resonant level model coupled to two auxiliary spin degrees of freedom instead of one \cite{guerci_unbinding_2017}.

We conclude this section by commenting on the choice of the slave-spin representation. Other slave-particle approaches, such as the slave-boson mean-field method in the infinite-$U$ limit~\cite{schwab_andreev_1999,rozhkov_interactingimpurity_2000} cannot identify the singlet-doublet transition, although this can be corrected with a finite-$U$ version of the method~\cite{bergeret_josephson_2007}. However, the advantage is the discrete gauge symmetry of the slave-spins, as opposed to the continuous $U(1)$ symmetry of the bosonic version, so that the Hilbert space is doubled instead of infinitely enlarged. The slave-spin representation is therefore more minimal and the resulting gauge fluctuations more under control. For example, it is known that the Kondo effect in this language appears as a spontaneous breaking of the $Z_2$ symmetry, which is correctly captured by mean-field theory and not washed out by fluctuations.

\section{Mean-field theory} \label{sec:meanfield}

The simplest way to solve Eq.~(\ref{Htotal}) is by means of a mean-field approximation. In such an approach, the wavefunction of the ground state is assumed to be factorized into a fermionic part and a spin part. We therefore assume that the ground state wavefunction of $\mathcal{H}'$ is a product state
\begin{equation}
    \ket{\Psi}= \ket{\Psi_{F}} \otimes \ket{\Psi_S},
\end{equation}
where $\ket{\Psi_{F}}$ corresponds to the fermionic part and $\ket{\Psi_S}$ corresponds to the auxiliary spin part. The mean-field decoupling results in two effective Hamiltonians
\begin{subequations}
\begin{align}
       \mathcal{H}_{F} &=\expval{\mathcal{H}'}{\Psi_S} \nonumber\\
       &= \mathcal{H}_{\mathrm{BCS}}+ \sum_{jk\sigma}  \overline{V}_j \pqty{c^{\dagger}_{jk\sigma}d_{\sigma}+d^{\dagger}_{\sigma}c_{jk\sigma}}, \label{RLM}\\
      \mathcal{H}_S  &=\expval{\mathcal{H}'}{\Psi_{F}}=h\sigma^x -\frac{U}{4}\sigma^z.\label{TLS}
\end{align}
\end{subequations}
Equation~(\ref{RLM}) describes a non-interacting S-QD-S system with renormalized hopping amplitudes 
\begin{equation}
    \overline{V}_j=\expval{\sigma^x}{\Psi_S} V_j=m_x V_j,
\end{equation}
while Eq.~(\ref{TLS}) describes a spin-1/2 system in a field with components $(h,0,-U/4).$ The component $h$ is given by the expectation value of the hybridization energy between the QD and the reservoirs 
\begin{equation}\label{hdefinition}
  h=\expval{\sum_{jk\sigma} V_j \pqty{c^{\dagger}_{jk\sigma}d_{\sigma}+d^{\dagger}_{\sigma}c_{jk\sigma}}}{\Psi_{F}}.  
\end{equation}
The two equations (\ref{RLM}) and (\ref{TLS}) are coupled through the parameters $m_x$ and $h.$ Calculation of these two parameters will lead to a coupled set of equations which will have to be solved self-consistently.

\subsection{Quantum dot coupled to superconducting leads}

For the solution of the non-interacting S-QD-S Hamiltonian (\ref{RLM}), it is convenient to work in Nambu space. We therefore introduce the spinors $\psi^{\dagger}_{jk}=\mqty(c^{\dagger}_{jk\uparrow} & c_{j-k\downarrow})$ and $\psi^{\dagger}_{d}=\mqty(d^{\dagger}_{\uparrow} & d_{\downarrow}).$ The main object of interest is the dot Green's function. In Nambu space, we define the (imaginary time) dot Green's function as
\begin{equation}\label{greenfunctiondot}
    \begin{split}
     \mathcal{G}_{dd}(\tau)&=-\expval{\mathcal{T}_{\tau}\psi_d(\tau)\psi^{\dagger}_{d}(0)}\\
     &=-\pmqty{\expval*{\mathcal{T}_{\tau}d_\uparrow(\tau)d^{\dagger}_{\uparrow}(0)} & \expval*{\mathcal{T}_{\tau}d_\uparrow(\tau)d_\downarrow(0)} \\ & \\ \expval*{\mathcal{T}_{\tau}d^{\dagger}_\downarrow(\tau)d^{\dagger}_\uparrow(0)} & \expval*{\mathcal{T}_{\tau}d^{\dagger}_\downarrow(\tau)d_\downarrow(0)}},
    \end{split}
\end{equation}
where the expectation values are taken with respect to Eq.~(\ref{RLM}). Since this is a non-interacting problem, we can exactly calculate the Green's function in Eq.~(\ref{greenfunctiondot}). Using the equation-of-motion method, we find a set of equations
\begin{subequations}\label{eom}
\begin{align}
  &  \mathcal{G}_{jd}(i\omega) = g_j(i\omega) \overline{V}_j \tau_z \mathcal{G}_{dd}(i\omega),\\
  &  \mathcal{G}_{dd}(i\omega) =\bqty{i\omega\mathds{1}_2-\sum_j \overline{V}_j^2\tau_z g_j(i\omega) \tau_z}^{-1}.
\end{align}
\end{subequations}
We have defined the lead-dot Green's function as $\mathcal{G}_{jk,d}(\tau)=-\expval{\mathcal{T}_{\tau}\psi_{jk}(\tau)\psi^{\dagger}_{d}(0)},$ $\tau_i$ as the $i$th Pauli matrix, and
\begin{equation}\label{reservoir}
g_j(i\omega)= \frac{-\pi \rho_0}{\sqrt{\Delta^2-(i\omega)^2}} \mqty(i\omega & \Delta e^{i\phi_j}\\ \Delta e^{-i\phi_j} & i\omega),
\end{equation}
as the Green's function of the isolated and semi-infinite $j$th superconductor written as a function of the fermionic Matsubara frequencies $i\omega.$ In calculating Eq.~(\ref{reservoir}) we have taken the wide-band limit, assuming a constant density of states $\rho_0$ in the normal state of the superconductor. We find
\begin{equation}
\mathcal{G}^{-1}_{dd}(i\omega)=i\omega \mathds{1}_2+\sum_j \frac{\overline{\Gamma}_j}{\sqrt{\Delta^2+\omega^2}} \mqty(i\omega & -\Delta e^{i\phi_j}\\ -\Delta e^{-i\phi_j} & i\omega),
\end{equation}
where $\overline{\Gamma}_j=\pi\rho_0\overline{V}_j^2$ is the hybridization with lead $j$. For convenience, we will consider only the case of symmetric couplings $\Gamma_L=\Gamma_R.$ In fact, we can do this without loss of generality since it has been shown that the properties of an asymmetric system can be related to those of the symmetric system \cite{kadlecova_quantum_2017}. Defining $\overline{\Gamma}_L=\overline{\Gamma}_R=\overline{\Gamma}/2$
we get
\begin{equation}\label{gdd}
\mathcal{G}_{dd}(i\omega) =\frac{1}{D(i\omega)}\mqty(i\omega \bqty{1+\alpha(i\omega)} & \alpha(i\omega)\Delta\cos\phi\\ \alpha(i\omega)\Delta\cos\phi & i\omega \bqty{1+\alpha(i\omega)}),
\end{equation}
where $D(i\omega)$ is the determinant of $\mathcal{G}_{dd}(i\omega)$ and 
\begin{equation}
  \alpha(i\omega)\equiv\frac{\overline{\Gamma}}{\sqrt{\Delta^2+\omega^2}}=\frac{2\pi\rho_0 \overline{V}^2}{\sqrt{\Delta^2+\omega^2}}.  
\end{equation}
We have denoted $\overline{\Gamma}=2\pi\rho_0 \overline{V}^2$ and $\Gamma=2\pi\rho_0 V^2,$ so that $\overline{\Gamma}=\Gamma m_x^2.$

\subsection{Self-consistent mean-field equations}
The renormalization parameter $m_x$ can be calculated from the solution of the Hamiltonian (\ref{TLS}), from which we get (for details see Appendix \ref{app:mf})
\begin{equation}\label{magnetization}
    m_x=-\frac{h}{E}, \qq{where} E=\sqrt{h^2+(U/4)^2}.
\end{equation}
In turn, the hybridization energy $h$ can be calculated as a function of the lead-dot Green's function,
\begin{equation}
\begin{split}
    h&=\frac{2V}{\beta}\sum_{i\omega} \sum_j \Re\bqty{\mathcal{G}_{jd}^{11}(i\omega)-\mathcal{G}_{jd}^{22}(i\omega)}e^{i\omega 0^+}\\
    &=\frac{4\overline{\Gamma}}{\beta m_x}\Re\sum_{i\omega}\frac{1}{D(i\omega)}\bqty{\frac{\overline{\Gamma}\Delta^2\cos^2\phi}{\Delta^2+\omega^2}+\frac{\omega^2(1+\alpha(i\omega))}{\sqrt{\Delta^2+\omega^2}}},\label{hmatsubara}
\end{split}
\end{equation}
The two equations (\ref{magnetization}) and (\ref{hmatsubara}) have to be solved self-consistently. At zero temperature, the sum over the Matsubara frequencies  of Eq.~(\ref{hmatsubara}) becomes an integral on the real axis
\begin{equation}\label{hintegral}
    h=-\frac{4}{m_x}\int_{-W}^{+W} \frac{\dd{\omega}}{2\pi} \frac{\overline{\Gamma}^2_\phi \Delta^2 +\overline{\Gamma}\omega^2 (\overline{\Gamma}+\sqrt{\omega^2+\Delta^2})}{\overline{\Gamma}^2_\phi\Delta^2 +\omega^2 (\overline{\Gamma}+\sqrt{\omega^2+\Delta^2})^2},
\end{equation}
where $\overline{\Gamma}_\phi=\overline{\Gamma}\cos\phi.$ We can solve this integral analytically in certain limits. Following \cite{schwab_andreev_1999}, we make the approximations
\begin{equation}
\frac{\omega}{\sqrt{\omega^2+\Delta^2}}=
    \begin{cases}
        1 &\omega>\Delta \\
        0 &\omega<\Delta
    \end{cases},
\end{equation}
and
\begin{equation}
\frac{\Delta}{\sqrt{\omega^2+\Delta^2}}=
    \begin{cases}
        0 &\omega>\Delta \\
        1 &\omega<\Delta
    \end{cases}.
\end{equation}
Then, at zero temperature, there are two contributions to the calculation of the hybridization energy $h,$
\begin{equation}\label{hanalytical}
\begin{split}
     h&=-\frac{4 m_x\Gamma}{\pi}\pqty{\int_0^\Delta \dd{\omega}\frac{\overline{\Gamma}\cos^2\phi}{\omega^2+\overline{\Gamma}_\phi^2}+\int_\Delta^W \dd{\omega}\frac{1}{\omega+\overline{\Gamma}}}\\
      &=-\frac{4m_x\Gamma_\phi}{\pi}\arctan(\frac{\Delta}{\overline{\Gamma}_\phi})-\frac{4 m_x\Gamma}{\pi}\ln(\frac{W+\overline{\Gamma}}{\Delta+\overline{\Gamma}}).
\end{split}
\end{equation}
\begin{figure}
    \centering
    \includegraphics[width=0.4\textwidth]{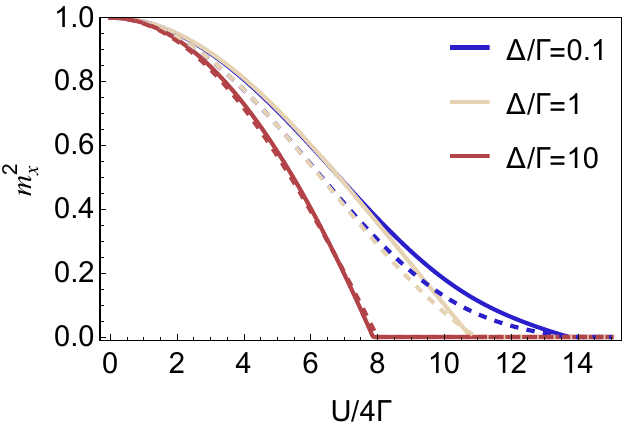}
    \caption{
   \justifying Solution to the self-consistent mean-field equations for various values of the superconducting gap, obtained from Eq.~(\ref{hintegral}) taking into account a finite bandwidth correction (dashed lines) and from the approximate analytical solution Eq.~(\ref{hanalytical}) (continuous lines). Parameters used: $\Gamma=0.001W,$ $W=100,$ $\phi=0.$}
    \label{fig:mx_comparison}
\end{figure}
The first term dominates when $\Delta\gg \Gamma,$ becoming exact at the superconducting atomic limit, while the second term dominates in the opposite limit $\Delta\ll \Gamma.$ In Fig.~\ref{fig:mx_comparison}, we show the renormalization parameter $m_x^2$ as a function of the interaction, for various values of the gap and compare $m_x^2$ calculated from Eq.~(\ref{hanalytical}) to the one calculated from Eq.~(\ref{hintegral}). Moreover, a finite bandwidth correction~\footnote{A finite bandwidth $W$ of the superconducting leads leads to a modification of the bare Green's functions in Eq.~(\ref{reservoir}), which have to be multiplied by a factor of $(2/\pi)\arctan(W/\sqrt{\Delta^2-(i\omega)^2}).$ This correspondingly leads to a modified coupling strength $\Gamma_w=(2\Gamma/\pi)\arctan(W/\sqrt{\Delta^2-(i\omega)^2})$} is taken into account in the calculation of Eq.~(\ref{hintegral}). The agreement between the two solutions is good, provided that the parameters of the problem $U,\Gamma,\Delta$ remain reasonably smaller than $W,$ although in the intermediate region, $\Delta\sim \Gamma,$ the approximate analytical solution slightly underestimates the effect of $U.$ However, since there is no significant quantitative difference and using an analytical equation is faster computationally, we use Eq.~(\ref{hanalytical}) for the calculation of $m_x$ in the rest of the paper and do not add any finite bandwidth correction to the Green's functions.

\subsection{Mean-field phase diagram}
\begin{figure}
	\centering
	\includegraphics[width=0.4\textwidth]{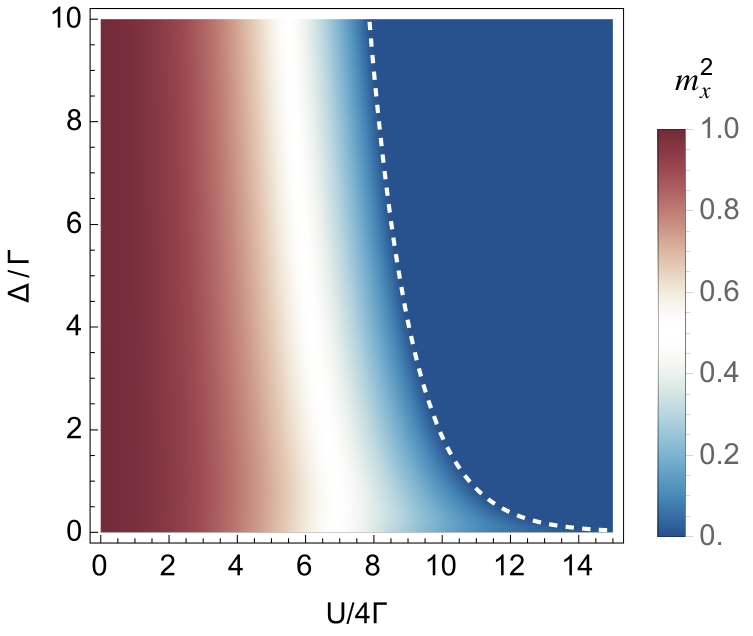}
	\caption{ \justifying Phase diagram of the superconducting AIM at half-filling. The phase difference is $\phi=0.$ The blue color corresponds to the doublet ground state, where the Kondo effect is suppressed. $\Gamma=0.001W, W=100.$ The white dashed line is the transition line separating the two phases, given by Eq.~(\ref{ucritical}).}
     \label{fig:phase diagram}
\end{figure}

The renormalized coupling $\overline{\Gamma}=\Gamma m_x^2$ can be interpreted as the Kondo temperature $T_K$ \cite{baruselli_subohmic_2012,guerci_unbinding_2017}. In the slave-spin mean-field theory, $m_x^2$ is finite in the singlet phase and zero in the doublet phase where the Kondo effect is fully suppressed. Within the slave-spin representation, the Kondo effect is therefore associated with a spontaneous breaking of the $Z_2$ \emph{gauge} symmetry associated to the auxiliary degree of freedom. Importantly, this has been shown to be a true feature of the problem~\cite{baruselli_subohmic_2012,zitko_2015} and not a mean-field artifact. We therefore use the quantity $m_x^2$ as an order parameter in order to construct a phase diagram for the superconducting AIM, which we present in Fig.~\ref{fig:phase diagram}.

In the absence of superconductivity, $\Delta=0,$ one expects to recover the known slave-boson mean-field expression for the Kondo temperature in the presence of normal leads~\cite{baruselli_subohmic_2012}. At zero temperature, the average hybridization energy simplifies to
\begin{equation}
      h =-\frac{4\overline{\Gamma}}{\pi m_x}\int_{0}^{W}\frac{\dd{\omega}}{\omega+\overline{\Gamma}} \simeq-\frac{4 m_x\Gamma}{\pi} \ln(\frac{W}{m_x^2\Gamma}).
\end{equation}
The non-trivial solution of the self-consistent equations in the absence of a superconducting gap and at $U\gg\Gamma$ is
\begin{equation}\label{eq:TK}
   \overline{\Gamma}= \Gamma m_x^2\simeq W e^{-\pi U/16\Gamma},
\end{equation}
and there is always a trivial solution $m_x=0.$ This energy scale plays the role of an effective Kondo temperature for the model. We note that the exponential scaling at large $U/\Gamma$ is captured well as compared to the exact Bethe ansatz result~\cite{meden_anderson_2019}, although the exponential factor $-\pi U/16\Gamma$ is two times smaller than the Bethe ansatz one. As a consequence, mean-field  overestimates the value of the Kondo temperature.

In presence of a small superconducting gap $\Delta\ll \Gamma,$ the renormalized hybridization remains finite, up to a critical value of interaction $U_c,$ above which it vanishes via a continuous quantum phase transition from the Kondo screened phase to an unscreened local moment phase, as we show in Fig.~\ref{fig:phase diagram}. This transition is also a transition from a singlet phase (due to the Kondo effect) to a magnetic doublet state represented by the free impurity decoupled from the bath continuum. At the critical value of the interaction $U_c(\Delta)$ the parameter $m_x^2$ becomes zero. In the limit $m_x\to 0,$ we linearize the hybridization energy $h$ in Eq.~(\ref{hanalytical}). By solving the self-consistent Eq.~(\ref{magnetization}) we find the singlet-doublet transition boundary
\begin{equation}\label{ucritical}
    U_c=\frac{16\Gamma}{\pi}\bqty{\frac{\pi}{2}\cos\phi+\ln(\frac{W}{\Delta})}.
\end{equation}

In Fig.~\ref{fig:phase diagram} we plot a phase diagram of the model within the mean-field theory by plotting the order parameter $\overline{\Gamma}/\Gamma=m_x^2$ as a function of $\Delta/\Gamma$ and $U/\Gamma$. 

\begin{figure}
    \centering
    \includegraphics[width=0.4\textwidth]{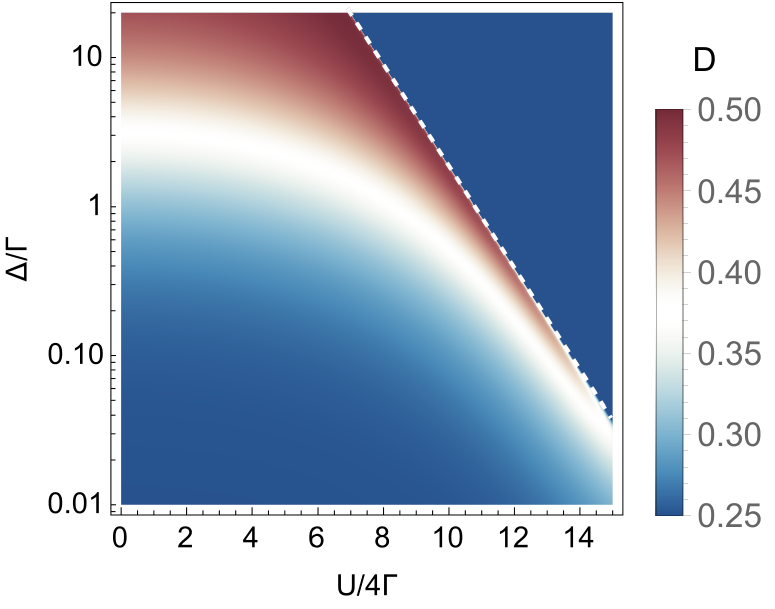}
    \caption{\justifying Double occupancy of the quantum dot, $D=\expval{n_\uparrow n_\downarrow}.$ The vertical axis is in logarithmic scale and parameter values are the same as in Fig.~\ref{fig:phase diagram}. The white dashed line indicates the singlet-doublet transition line.}
    \label{fig:double_occupancy}
\end{figure}

As the superconducting gap is further increased, new physics is expected to come into play. Indeed, $\Delta$ brings into the system pairs of electrons with opposite spin, therefore changing the charge structure of the Kondo state but not its spin. Within the singlet phase at $U<U_c(\Delta)$ we therefore expect a crossover from a Kondo singlet to a BCS singlet made of superposition of doublons and holons. To probe this crossover, it is useful to compute the fraction of doublons on the impurity site. This amounts to calculating the double occupancy on the dot $D=\expval{n_\uparrow n_\downarrow}.$ In the mean-field theory $D=\expval*{d^\dagger_\uparrow d_\uparrow}\expval*{d^\dagger_\downarrow d_\downarrow}-\expval*{d^\dagger_\uparrow d^\dagger_\downarrow}\expval*{d_\uparrow d_\downarrow}.$ The result is plotted in Fig.~\ref{fig:double_occupancy}. This quantity is equal to $1/4$ in the non-interacting, non-superconducting limit $U=0, \Delta=0.$ Within the singlet phase, it increases up to $1/2$ at large $\Delta$ and moderate interactions, where a crossover from Kondo to BCS-like state is clearly visible. For large values of the interaction and a finite gap, the mean-field is unable to capture the $\pi$ phase, falsely giving a nonzero double occupancy at large interaction values. We note that the mean-field decoupling fails to properly describe the regime at large $\Delta$, where entanglement (and correlations) between the dot and the auxiliary two-level system is crucial to account for the right physics. This is more clearly seen in the limit $\Delta\rightarrow\infty$.

In the large gap limit, it is well known that the physics is in fact described by an effective dot-only Hamiltonian containing the exact Coulomb interaction and an effective pairing term~\cite{PhysRevB.79.224521}. Such a Hamiltonian describes a level crossing between a singlet and doublet ground state at a critical value of the interaction $U_c=2\Gamma_\phi.$ In this limit, the summation over the Matsubara frequencies in Eq.~(\ref{hintegral}) can be easily performed with contour integration. At zero temperature,
\begin{equation}
    h=-\frac{4\overline{\Gamma}_\phi^2}{\beta m_x}\sum_{i\omega} \frac{1}{\omega^2+\overline{\Gamma}^2_\phi}=-2m_x\Gamma_\phi.
\end{equation}
Then, the solution of the self-consistent equations is
\begin{equation}
    m_x^2=1-\pqty{\frac{U}{8\Gamma_\phi}}^2,
\end{equation}
if $U\leq 8\Gamma_\phi,$ and $m_x^2=0$ otherwise. The critical value $U_c=8\Gamma_\phi$ is four times bigger than the value expected in the large gap limit~\cite{meng_selfconsistent_2009}. Furthermore, while $U_c$ signals again a singlet-doublet transition, this appears to be concomitant in mean-field theory with a transition in the pseudo-spin magnetization, which is, however, absent in the exact solution, as we discuss in Appendix~\ref{sec:largegap}.

\subsection{Comparison with Previous Literature}
The slave-spin mean-field result for the transition boundary is
\begin{equation}\label{dcritical}
    \Delta_c=W e^{\frac{\pi}{2}\cos\phi} e^{-\pi U/16\Gamma}.
\end{equation}
which is equivalent to the condition $\Delta=e^{\frac{\pi}{2}\cos\phi}\overline{\Gamma}(0),$ with $\overline{\Gamma}(0)$ being the Kondo temperature in the slave-spin mean-field theory of the normal state.

\begin{figure}
    \centering
    \includegraphics[width=0.4\textwidth]{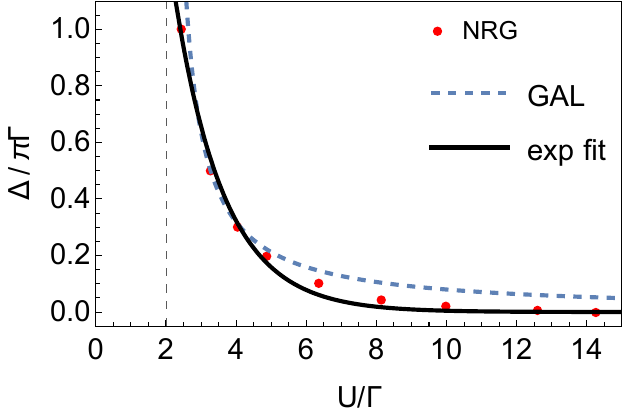}
    \caption{\justifying{Singlet-doublet transition line and comparison to known results. The red dots are NRG data taken graphically from Ref.~\cite{bauer_spectral_2007} and fitted using a rescaled Eq.~\ref{dcritical} (black line). The blue dashed line is the generalized atomic limit result and the vertical black dashed line is the large gap limit result. $\pi\Gamma=0.2 W.$}}
    \label{fig:comparison}
\end{figure}

For small values of $\Delta/\Gamma,$ the transition is expected to occur when $T_K/\Delta\sim 0.3$ with the Kondo temperature in the normal state $T_K$ defined as $T_K=0.182 \sqrt{\frac{8\Gamma U}{\pi}}e^{-\pi U/8\Gamma}$ \cite{bauer_spectral_2007}. We see that Eq.~(\ref{dcritical}) captures the correct exponential behavior, but gives the wrong exponential factor and overestimates the prefactor. In order to get a clearer picture, we compare our analytic result for the transition boundary Eq.~(\ref{dcritical}) with the NRG results (obtained graphically) of reference \cite{bauer_spectral_2007}. We find that we can fit the NRG data with an exponential of the form $\Delta_c= (W/\alpha) e^{\pi/2} e^{-\pi U/\beta \Gamma},$ where $\alpha=4.34,$ and $\beta=4.42.$ This means that the correct behavior can be recovered by a suitable rescaling of parameters. In Fig.~\ref{fig:comparison} we plot the transition line as obtained from Eq.~\ref{dcritical} but with rescaled $\Gamma\to \beta\Gamma/16$ and $W\to W/\alpha.$ The resulting curve (solid black line) compares reasonably well with the NRG data (red dots). We also compare with the large gap limit result $U/2=\Gamma \cos\phi$ (vertical dashed line), as well as with the generalized atomic limit (GAL) result (blue dashed line) which is obtained from a perturbation theory in the interaction $U$. The GAL resembles the large gap limit, but permits a correction for a finite superconducting gap and predicts that the transition line is given by $U/2=(1+\frac{\Gamma}{\Delta})\Gamma \cos\phi$ \cite{zonda_perturbation,zonda_perturbation_2016,zonda_generalized_2023}.

\subsection{Physical Green's function}

An important quantity to characterize the physics of our model is the dot Green's function, in particular the diagonal and off diagonal component in Nambu space defined as
\begin{equation}
\begin{split}
     G^{11}_{ff}(\tau)&=-\expval*{\mathcal{T}_{\tau} f_{\uparrow}(\tau)f_{\uparrow}^{\dagger}(0)},\\
      G^{12}_{ff}(\tau)&=-\expval{\mathcal{T}_{\tau}f_{\uparrow}(\tau)f_{\downarrow}(0)}.
\end{split}
\end{equation}
In the mean-field theory the Green's function factorizes into
\begin{equation}\label{greenfunctionphysical}
    G^{ab}_{ff}(\tau)= B_{xx}(\tau)\mathcal{G}_{dd}^{ab}(\tau),
\end{equation}
where
\begin{equation}
    B_{xx}(\tau)=\expval{\mathcal{T}_{\tau} \sigma^x(\tau)\sigma^x(0)}_{S}
\end{equation}
is a spin-spin correlation function and $\mathcal{G}_{dd}$ is given in Eq.~(\ref{greenfunctiondot}). Fourier transform turns the product in Eq.~(\ref{greenfunctionphysical}) into a convolution, so that
\begin{equation}\label{convolution}
     G_{ff}(i\omega)=\frac{1}{\beta}\sum_{i\varepsilon}B_{xx}(i\omega-i\varepsilon)\mathcal{G}_{dd}(i\varepsilon),
\end{equation}
where $i\omega, i\varepsilon$ are fermionic Matsubara frequencies. The $\sigma^x$ autocorrelation function, calculated in Appendix~\ref{app:bxx}, is equal to
\begin{equation}\label{spincor}
    B_{xx}(i\Omega)=m_x^2 \delta(i\Omega)+\Pi_{xx}(i\Omega),
\end{equation}
where $i\Omega$ is a bosonic Matsubara frequency. $B_{xx}$ is composed of a coherent contribution proportional to $m_x^2$ as well as an incoherent part which describes the high-energy behavior,
\begin{equation}\label{inc_mf}
\begin{split}
       \Pi_{xx}(i\Omega)&=(1-m_x^2)\tanh(\beta E)\\ &\times\pqty{\frac{1}{i\Omega+2E}-\frac{1}{i\Omega-2E}}.
\end{split}
\end{equation}
With these definitions, the Green's function of the $f-$fermions is given by
\begin{equation}\label{eq:physicalGreensmatsubara}
      G_{ff}(i\omega)=m_x^2\mathcal{G}_{dd}(i\omega)+\frac{1}{\beta}\sum_{i\varepsilon}\Pi_{xx}(i\omega-i\varepsilon)\mathcal{G}_{dd}(i\varepsilon).
\end{equation}
This expression can be rewritten using the spectral representations of the fermionic and spin Green's functions. Introducing the spectral representations
\begin{equation}
    \begin{split}
      &\Pi_{xx}(i\Omega)=\int\dd{x} \frac{\rho_s(x)}{i\Omega-x} \\
      &\mathcal{G}_{dd}(i\omega)=\int\dd{y}\frac{\rho_d(y)}{i\omega-y}
    \end{split}
\end{equation}
Eq.~(\ref{eq:physicalGreensmatsubara}) becomes
\begin{equation}\label{eq:physicalGreen}
\begin{split}
    G_{ff}&(i\omega)=m_x^2\mathcal{G}_{dd}(i\omega)\\ &+\iint\dd{x}\dd{y}\frac{\rho_s(x)\rho_d(y)}{i\omega-x-y}\bqty{n_F(y)+n_B(-x)}  
\end{split}
\end{equation}
where $n_F(x)$ is the Fermi-Dirac distribution and $n_B(x)$ is the Bose-Einstein distribution. 

\subsection{Spectral function and Andreev bound states}

Analytical continuation \footnote{One has to be careful with the analytical continuation $i\omega\to\omega+i\eta.$ Essentially, the square-root function should be continued as follows:
\begin{equation}
   \frac{1}{\sqrt{\Delta^2+\omega^2}}\to
    \begin{cases}
    \frac{1}{\sqrt{\Delta^2-\omega^2}} & \abs{\omega}<\Delta \\    
    \frac{i\mathrm{sign}(\omega)}{\sqrt{\omega^2-\Delta^2}} & \abs{\omega}>\Delta
    \end{cases}
\end{equation}
} of the physical Green's function to real frequencies $i\omega\to\omega+i\eta$ gives access to the spectral function, which we define as
\begin{equation}\label{eq:spectralfunction}
\begin{split}
    \mathcal{A}(\omega)&=-\frac{1}{\pi}\Im G^{11}_{ff}(\omega+i\eta)\\
    &=m_x^2\rho^{11}_{d}(\omega)+\int\dd{x}\rho_s(x)\rho^{11}_d(\omega-x)\\&\times\bqty{n_F(\omega-x)+n_B(-x)}.
\end{split}
\end{equation}
At zero temperature, the mean-field spin spectral function has a three-peak structure, so that we find
\begin{equation}
\begin{split}
   \mathcal{A}(\omega)&=m_x^2 \rho^{11}_d(\omega)\\ &+(1-m_x^2)\bqty{\rho^{11}_d(\omega+2E)+\rho^{11}_d(\omega-2E)}.  
\end{split}
\end{equation}
\begin{figure}
	\centering
	\begin{subfigure}{0.95\linewidth}
		\centering
		\includegraphics[width=\textwidth]{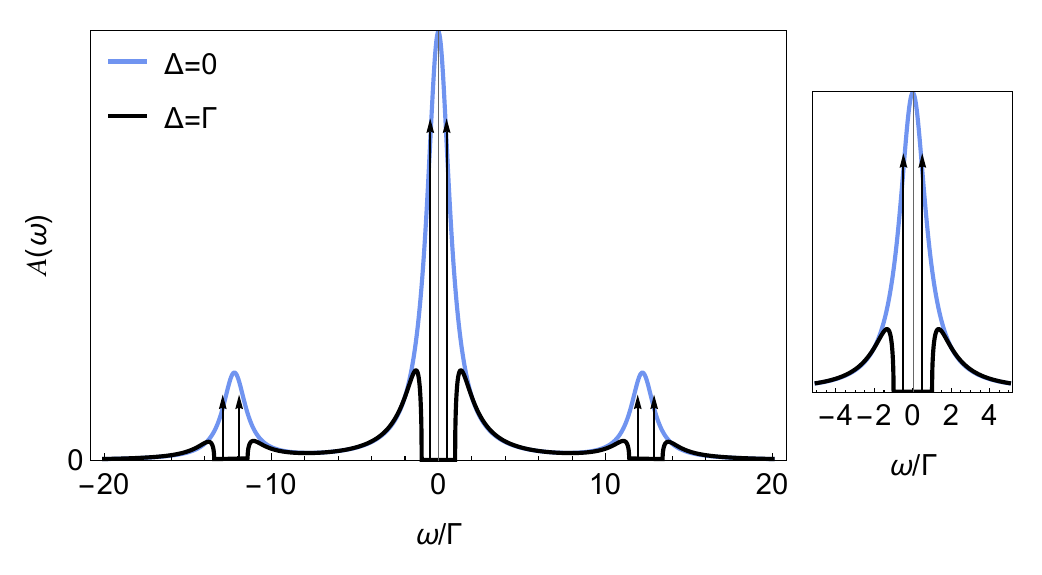}
  		\caption{}
    \label{spectral_mf}
	\end{subfigure}
	\hfill
	\begin{subfigure}{0.8\linewidth}
		\centering
		\includegraphics[width=\textwidth]{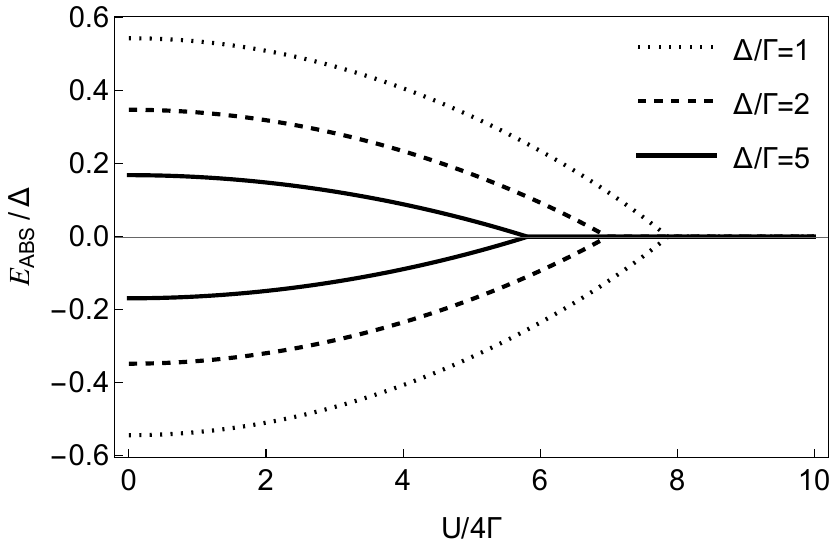}
  		\caption{}
    \label{ABS}
	\end{subfigure}
	\caption{\justifying (a) Spectral function for $U=10\Gamma$ and $\Gamma=0.01W.$ The blue (black) line corresponds to the absence (presence) of superconductivity. The right panel shows a zoom in the low energy region. The position of the Andreev bound states is represented by arrows whose height corresponds to the spectral weight. (b) Andreev bound state energies in the subgap region as a function of the interaction $U$ for various values of the gap and $\Gamma=0.01W.$}
\end{figure}
The above equation gives the correct description for the low-energy window, but displays some unphysical features at high-energies. This is illustrated in Fig.~\ref{spectral_mf}. In the absence of superconductivity, a three-peak structure appears. The central peak can be interpreted as the Kondo resonance, while the high-energy peaks can be understood as the Hubbard bands. The first unphysical feature of the mean-field result is that all three peaks are being broadened with the same width, which here corresponds to $\Gamma m_x^2.$ This leads to an unphysical narrowing of the Hubbard bands as $U$ becomes large. At finite $\Delta,$ the Kondo resonance does not fully form, as expected, and is suppressed inside the gap where ABS form. However, we also observe the formation of discrete states in the high-energy region. This unphysical behavior motivates us to go beyond the mean-field result. In the next section, we will show that including RPA corrections in the spin autocorrelation function is sufficient to retrieve the correct high-energy behavior of the spectral function.

Figure~\ref{ABS} shows the dependence of the energies of the subgap states $E_{\mathrm{ABS}}$ on the interaction. The ABS energies are found by setting the determinant of $\mathcal{G}_{dd}(\omega)$ to zero, which amounts to finding the solutions to the following equation
\begin{equation}\label{eq:ABS}
    \omega^2(\overline{\Gamma}+\sqrt{\Delta^2-\omega^2})^2-(\overline{\Gamma}\Delta\cos\phi)^2=0.
\end{equation}
The ABS start from their non-interacting value and progressively approach zero with increasing interaction, reaching zero at the boundary of the $0-\pi$ transition. It is, however, expected that the ABS cross at the $0-\pi$ transition. Indeed, the ABS represent excitations that change the particle number by 1, i.e. transitions between the spin doublet (odd parity) states  and the spin singlet (even parity) states \cite{PhysRevB.79.224521}. Since at the $0-\pi$ transition the two states (singlet and doublet) are degenerate, one expects that the ABS energy at the transition is zero. We see that the slave-spin method is unable to access the behavior in the $\pi$ phase, while for the $0$ phase, the mean-field description of the ABS is qualitatively sufficient. This is because in the $\pi$ phase, the slave-spin mean-field decoupling erroneously describes the dot as a resonant level model decoupled from the leads.

\section{RPA-Corrections to Mean-Field Theory}
\label{sec:beyond}
In this section we go beyond the simple mean-field approach described so far, by including fluctuations corrections to the spin dynamics within a Random Phase Approximation approach (RPA)~\cite{guerci_transport_2019}, which we generalize here to the case of a superconducting bath. As we will show below, this will allow to correct the high-frequency features of the mean-field spectrum that displayed Hubbard bands with a gap and a width controlled by the renormalized hybridization $\overline{\Gamma}$, rather than the bare $\Gamma.$ For simplicity, the physical fermion Green's function will still be calculated using a mean-field decoupling between spins and  electrons, see Eq.~(\ref{greenfunctionphysical}) that we report here for simplicity
\begin{equation}\label{greenfunctionphysical2}
    G^{ab}_{ff}(\tau)= B_{xx}(\tau)\mathcal{G}_{dd}^{ab}(\tau)\,,
\end{equation}
with $a,b$ Nambu indices. Our goal is to compute the spin autocorrelation function $B_{xx}(\tau)$ including quantum fluctuations beyond mean-field. To formulate the RPA it is convenient to use Feynman diagram techniques. To this end, we rewrite the spin operators by introducing Abrikosov pseudofermions \cite{abrikosov_electron_1965,coleman,oppermann_application_1973}. The motivation behind this trick is that spin operators obey non-standard commutation rules since $\comm{\sigma^i}{\sigma^j}=2i\varepsilon_{ijk}\sigma^k,$ and Wick's theorem cannot be applied. The components of the spin operators are rewritten as
\begin{equation}
    \sigma^i=\sum_{\alpha\beta}\psi^{\dagger}_\alpha \sigma^i_{\alpha\beta}\psi_\beta,
\end{equation}
where the operators $\psi$ are fermionic: $\acomm{\psi^{\dagger}_\alpha}{\psi_\beta}=\delta_{\alpha \beta},$ and $\acomm{\psi_\alpha}{\psi_\beta}=0.$ For a spin-1/2 the indices $\alpha,\beta$ can only take two values, $\alpha,\beta=\pm.$ Details of the calculations are found in Appendix \ref{app:mf}. 
\begin{figure}
    \centering
    \includegraphics[width=0.4\textwidth]{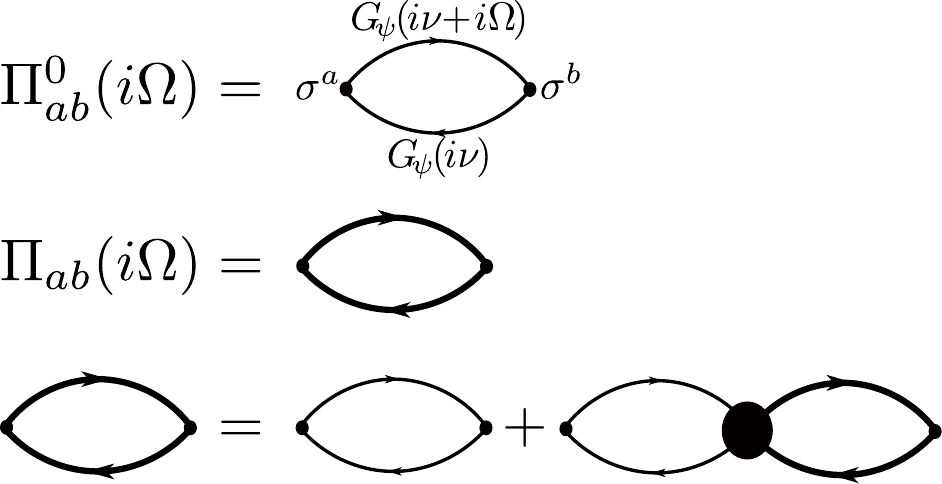}
    \caption{\justifying Random phase approximation (RPA) for the spin autocorrelation function.}
    \label{fig:diagrams}
\end{figure}

The propagator for the $\psi$-fermions can be defined as 
\begin{equation}
    G^{\alpha\beta}_{\psi}(\tau)=-\expval{\mathcal{T}_\tau \psi_\alpha(\tau)\psi^{\dagger}_\beta(0)}.
\end{equation}
In terms of pseudofermions we can rewrite the slave-spin Hamiltonian (\ref{Htotal}) as
\begin{equation}\label{eqn:Hprime}
\begin{split}
  \mathcal{H}'&=\mathcal{H}_{\mathrm{BCS}}-\frac{U}{4}\sum_{\alpha\beta}\psi^{\dagger}_\alpha \sigma^z_{\alpha\beta}\psi_\beta \\
  &+ \sum_{jk\sigma}\sum_{\alpha\beta}V \pqty{c^{\dagger}_{jk\sigma}\psi^{\dagger}_\alpha \sigma^x_{\alpha\beta}\psi_\beta d_\sigma +\mathrm{H.c.}}.
\end{split}  
\end{equation}
If the pseudofermions are free ($V=0$), the bare propagator is
\begin{equation}
G^0_\psi(i\omega)=(i\omega+U\sigma^z/4)^{-1},
\end{equation}
and the bare $\sigma^x$ autocorrelation function can be calculated,
\begin{equation}
\begin{split}
    B^0_{xx}(i\Omega)&=m_x^2\delta(i\Omega)\\
    &-\frac{1}{\beta}\sum_{i\nu}\Tr_{\sigma}\bqty{\sigma^x G^0_\psi(i\Omega+i\nu)\sigma^x G^0_\psi(i\nu)},  
\end{split}
\end{equation}
where $\Omega$ are bosonic and $\nu$ are fermionic Matsubara frequencies. A finite hybridization $V\neq0$ in Eq.~(\ref{eqn:Hprime}) induces an interaction vertex between the fermions in the bath, the dot and the pseudo-fermions, leading to two main effects. First, the pseudo-fermion Green's function acquires now a self-energy correction $\Sigma_{\psi}$
\begin{equation}
  G_\psi=G_\psi^0+G_\psi^0\Sigma_\psi G_\psi\,.
\end{equation}
This self-energy can be computed to lowest order using the Hartree-Fock approximation to obtain $\Sigma_\psi=\sigma^x h,$ where $h$ is the average hybridization defined in Eq.~(\ref{hdefinition}).

As a consequence, the (incoherent part of the) spin auto-correlation function also gets renormalized into an RPA bubble with internal lines given by $G_\psi$, i.e.
\begin{equation}\label{eq:pirr}
\begin{split}
        \Pi^{\mathrm{irr}}_{RPA}&=\Pi^0_{xx}\\ &=-\frac{1}{\beta}\sum_{i\nu}\Tr_{\sigma}\bqty{\sigma^x G_\psi(i\Omega+i\nu)\sigma^x G_\psi(i\nu)}.
\end{split}
\end{equation}
As we show in Appendix \ref{app:bxx}, this expression reproduces the mean-field result given in Eq.~(\ref{inc_mf}). To go beyond mean-field one needs to account for vertex corrections in computing the polarization bubble $\Pi_{xx}$, consistently with the Hartree-Fock approximation to make sure to be conserving~\cite{guerci_transport_2019}.
This gives rise to a Dyson equation for the incoherent contribution to the spin autocorrelation, which reads
\begin{equation}
    \Pi_{xx}=\Pi^0_{xx}+\Pi^0_{xx}\Sigma_{xx}\Pi_{xx},
\end{equation}
where $\Pi^0_{xx}$ is given in Eq.~(\ref{eq:pirr}) while the self-energy term takes the form~\cite{guerci_transport_2019}
\begin{equation}
    \Sigma_{xx}(\tau)=\expval{\mathcal{V}}^2-\expval{\mathcal{T}_\tau \mathcal{V}(\tau)\mathcal{V}(0)},
\end{equation}
where $\mathcal{V}=\sum_{jk\sigma} V_j \pqty{c^{\dagger}_{jk\sigma}d_{\sigma}+\mathrm{H.c.}}$ is the tunneling operator. Fourier transformation gives
\begin{equation}\label{eq:sigmamatsubara}
\begin{split}
    \Sigma_{xx}(i\Omega)&=\frac{-V^2}{\beta}\sum_{jj'}\sum_{i\nu}\Tr_N\Bigg[\bigg(g_j(i\nu)\mathcal{G}_{dd}(i\Omega+i\nu)\\&+g_j(i\Omega+i\nu)\mathcal{G}_{dd}(i\nu)\bigg)\delta_{jj'}\\&+2\mathcal{G}_{jd}(i\Omega+i\nu)\mathcal{G}_{j'd}(i\nu)\Bigg],
\end{split}
\end{equation}
where $\Tr_N$ indicates a trace over the Nambu indices and the Green's functions are given in Eq.~(\ref{eom}) and (\ref{reservoir}). In Appendix \ref{app:rpa}, we show how the above expression can be analytically continued to real frequencies, $i\Omega\to \Omega+i\eta,$ and find its imaginary part to be
\begin{equation}\label{eq:sxx}
\begin{split}
\Im\Sigma_{xx}(\Omega\!+\!i\eta)&=2\pi V^2\int\dd{x}\bqty{n_F(x)\!-\!n_F(x\!+\!\Omega)}\\\ &\times \Tr_N\sum_j\rho_j(x)\rho_d(x+\Omega).
\end{split}
\end{equation}
The corresponding real part which gives a shift of the Hubbard bands is nonzero but small, and is ignored. In order to arrive at Eq.~(\ref{eq:sxx}) we also ignored the last term of Eq.~(\ref{eq:sigmamatsubara}) which is of order $\Gamma^2 m_x^2$ and therefore much smaller than the first two terms which are of order $\Gamma.$  At zero temperature, the self-energy given by Eq.~(\ref{eq:sxx}) has an extended gap, i.e. is non-zero if $\abs{\Omega}>\Delta+E_{\mathrm{ABS}},$ with $E_{\mathrm{ABS}}$ being the energy of the ABS. Moreover, we observe that at large frequencies $\abs{\Omega}\gg\Delta,$ the self-energy tends to a constant value given by $\Sigma_{xx}(\Omega)=2i\Gamma \mathrm{sign}(\Omega).$ Calculations can therefore be further simplified by taking the equivalent of a wide-band limit, which amounts to making the approximation $\Sigma_{xx}(\Omega)\simeq 2i\Gamma \mathrm{sign}(\Omega)$ when $\abs{\Omega}>\Delta+E_{\mathrm{ABS}}$ and $\Sigma_{xx}(\Omega)=0$ otherwise.

\subsection{Spectral function}
\begin{figure}
	\centering
	\begin{subfigure}{0.8\linewidth}
		\centering
		\includegraphics[width=\textwidth]{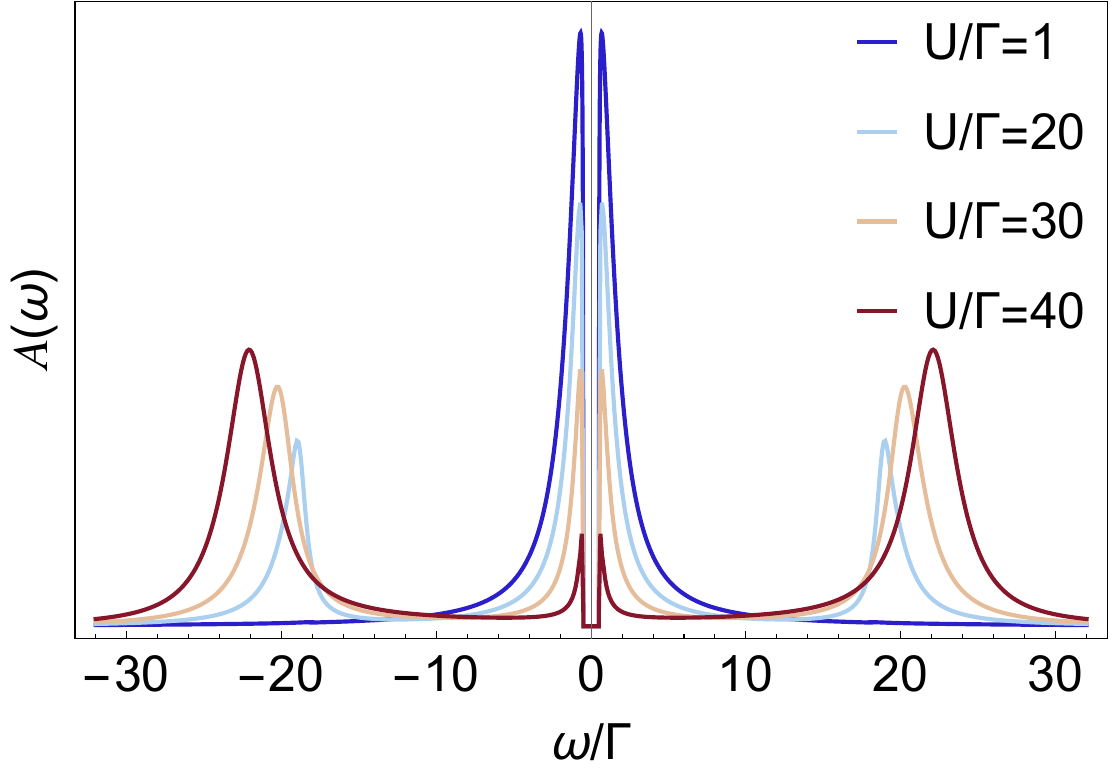}
  		\caption{}
   \label{fig:specRPA_a}
	\end{subfigure}
	\hfill
	\begin{subfigure}{0.8\linewidth}
		\centering
		\includegraphics[width=\textwidth]{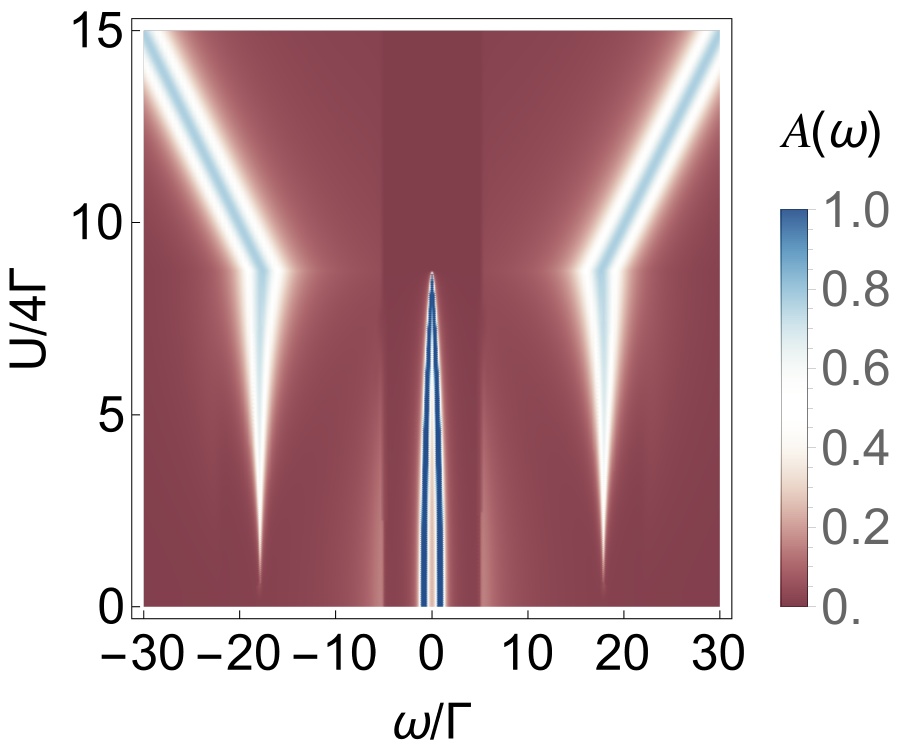}
  		\caption{}
    \label{fig:specRPA_b}
	\end{subfigure}
	\caption{\justifying (a) Dot spectral function after including RPA corrections, for various values of the interaction and $\Delta=\Gamma/2.$ Here the ABS are not shown for clarity. (b) Density plot of the spectral function as a function of the energy $\omega/\Gamma$ and the interaction $U/4\Gamma,$ for $\Delta=5\Gamma.$ The coupling strength is fixed to $\Gamma=0.001W$ in both cases.}
     \label{fig:specRPA}
\end{figure}
Using the RPA results for the spin correlation function, we can compute the dot spectral function, which we plot in Fig.~\ref{fig:specRPA_a} for various values of the interaction $U$ and a small gap $\Delta=\Gamma/2.$ The spectral function is calculated using Eq.~(\ref{eq:spectralfunction}), where now the spin spectral function $\rho_s(x)=-\frac{1}{\pi}\Im\Pi_{xx}(x+i\eta)$ is calculated within RPA. We see that the RPA corrects the unphysical results of the mean-field treatment, producing Hubbard bands centered around energies $\pm 2E=\pm \sqrt{4h^2+(U/2)^2}$ and with a width depending on the bare coupling $\Gamma.$ At $U\lesssim \Gamma$ the spectral function resembles the one found in the absence of superconductivity except for the opening of a gap and the formation of the ABS in the subgap region (in Fig.~\ref{fig:specRPA_a} the ABS are not shown for clarity). 

As the interaction increases, weight is transferred to the Hubbard bands and (what remains of) the Kondo resonance progressively shrinks. In Fig.~\ref{fig:specRPA_b} we plot the spectral function for a wide range of interaction values. We see that, similar to the mean-field treatment, the bound state energies reach zero at the $0-\pi$ transition beyond which they disappear, while the position of the Hubbard bands in the large interacting limit is correctly given by $\omega=\pm U/2.$ 

The RPA therefore succeeds in describing the whole energy spectrum in the singlet phase, but fails to describe the low-energy features in the doublet phase. While the Hubbard bands and a gap are correctly present in the $\pi$ phase, there is no proximity effect, meaning that the superconducting continua and the ABS disappear from the spectral function. Essentially, in the doublet phase the auxiliary fermions still become decoupled from the bath as within the previous mean-field treatment. However, RPA partly corrects this decoupling, because the auxiliary spins remain coupled to the leads through the self-energy $\Sigma_{xx}.$

\subsection{Josephson current}
The Josephson current can be calculated through the variation of the number operator in one of the leads
\begin{equation}\label{eq:josephson_current}
    J_j(t)=\dv{t}N_j(t)=i\comm{H}{N_j}=i\comm{H_{\mathrm{tun}}}{N_j},
\end{equation}
where $N_j=\sum_{k\sigma}c^{\dagger}_{jk\sigma}c_{jk\sigma}.$ Its expectation value, taken with respect to Hamiltonian (\ref{model}), can be written as
\begin{equation}
\begin{split}
    J_j&=iV\sum_{k\sigma}\bqty{\expval{f^{\dagger}_\sigma c_{jk\sigma}}-\expval{c^{\dagger}_{jk\sigma}f_\sigma}} \\
    &=-\frac{2V}{\beta}\sum_{i\omega} e^{i\omega 0^+} \Im\Tr_N G_{jf}(i\omega).
\end{split}
\end{equation}
We have defined $G_{jk,f}(\tau)=-\expval{\mathcal{T}_{\tau}\psi_{jk}(\tau)\psi^{\dagger}_{f}(0)}$ a lead-dot Green's function in Nambu space, involving the physical $f-$fermions. Using the equations-of-motion technique, we calculate
\begin{equation}
    G_{jf}(i\omega) = g_j(i\omega) V \tau_z G_{ff}(i\omega).
\end{equation}
The physical Green's function of the QD can be calculated using a mean-field decoupling, see Eq.~(\ref{eq:physicalGreensmatsubara}). Accordingly, the current for i.e. the left lead can be decomposed into a coherent term and an incoherent term which we note by $J_L=J_L^{coh}+ J_L^{inc}.$

The expression for the coherent term is the same as for the non-interacting Josephson current \cite{meden_anderson_2019}, but calculated with a renormalized coupling to the leads,
\begin{equation}
    J_L^{coh}=-\frac{1}{\beta}\sum_{i\omega}\frac{\overline{\Gamma}^2\Delta^2 \sin 2\phi}{D(i\omega)\pqty{\Delta^2+\omega^2}},
\end{equation}
 At $T=0^+,$ we find
\begin{equation}
      J_L^{coh}=\int_{-W}^{+W}\frac{\dd{\omega}}{2\pi}\frac{\overline{\Gamma}^2\Delta^2 \sin 2\phi}{(\overline{\Gamma}_\phi\Delta)^2 +\omega^2 (\overline{\Gamma}+\sqrt{\Delta^2+\omega^2})^2}.
\end{equation}
N.B. here $m_x$ is a function of $U$ and of $\phi.$ The incoherent term is given by
\begin{equation}
    J_L^{inc}=-\frac{2V^2}{\beta^2}\sum_{i\omega,i\varepsilon}\Im\Tr_N\bqty{g_L(i\omega)\tau_z\Pi_{xx}(i\omega\!-\!i\varepsilon)\mathcal{G}_{dd}(i\varepsilon)}.
\end{equation}
At $T=0^+,$
\begin{equation}
\begin{split}
     J_L^{inc}&=\frac{4\Gamma\Delta\sin\phi}{\pi}\iiint\dd{x}\dd{y}\dd{z} \theta(x)\theta(y)\theta(z-\Delta)\\
     &\times \frac{\rho_d^{12}(x)\rho_s(y)}{\sqrt{z^2\!-\!\Delta^2}(x\!+\!y\!+\!z)}.
\end{split}
\end{equation}
\begin{figure}
    \centering
    \includegraphics[width=0.4\textwidth]{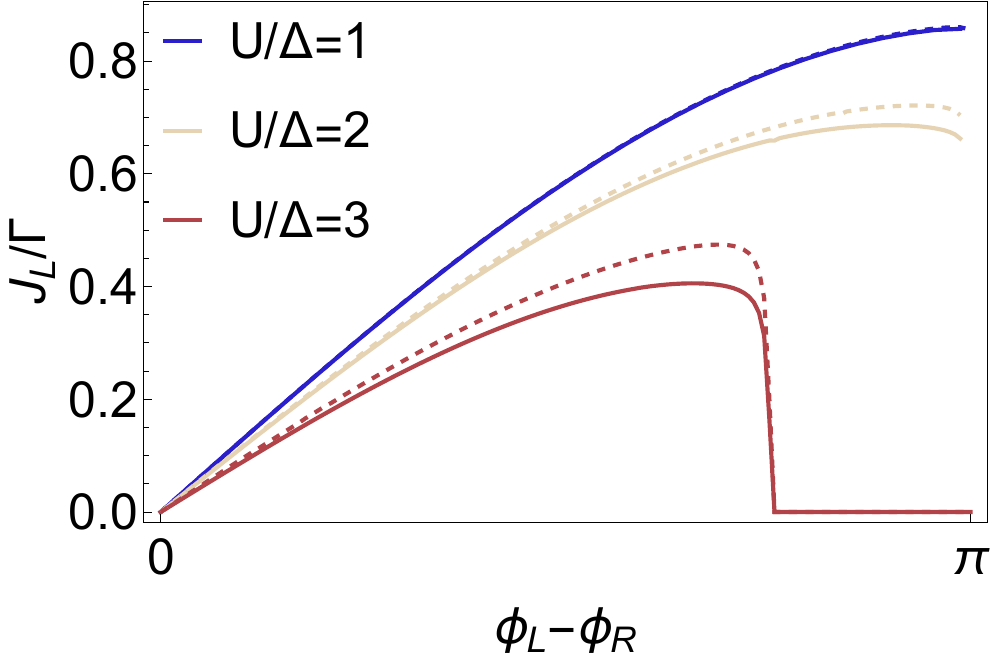}
    \caption{\justifying Josephson current for various interaction values and fixed gap $\Delta=10\Gamma,$ with $\Gamma=0.001W.$ The dashed lines indicate the mean-field result.}
    \label{josephson_current}
\end{figure}
Figure~\ref{josephson_current} shows the resulting Josephson current from the left lead as a function of the phase difference $\phi_L-\phi_R=2\phi.$ We only show the region $\phi_L-\phi_R\in \qty[0,\pi],$ with the reminder that the supercurrent is an odd function in the phase difference and has a period of $2\pi.$ Moreover, we compare the RPA result to the mean-field one. Since the Josephson current is mostly carried by the subgap states, the contribution from the high-energy features is expected to be small. RPA corrections do not change the mean-field result significantly for small values of $U/\Delta$ but tend to decrease the value of the critical current as the interactions grow stronger. 

The $0-\pi$ transition as a function of the phase is expressed as an abrupt jump of the supercurrent to zero. The rounding of the jump is a numerical artifact due to a finite imaginary part $\eta=10^{-6}W$ used in the definition of the spectral functions. As our method does not capture the proximity effect in the doublet phase, the off-diagonal component of the dot Green's function becomes zero there. As a consequence, both components of the Josephson current are always zero in the doublet phase, while the physical behavior would be a change of sign and a decrease in magnitude.

\subsection{Superconducting correlations}

\begin{figure}[t]
	\centering
	\begin{subfigure}{0.8\linewidth}
		\centering
		\includegraphics[width=\textwidth]{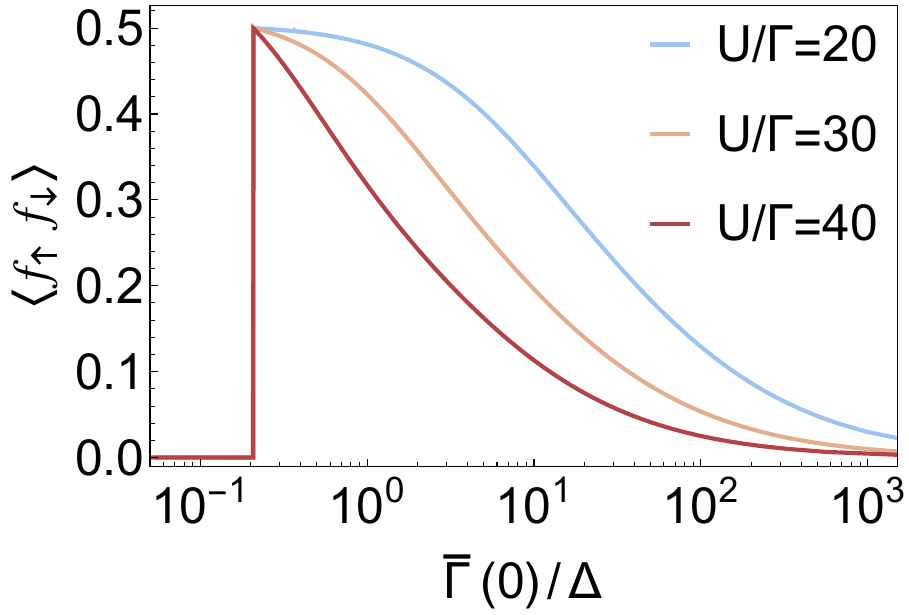}
  		\caption{}
   \label{fig:correlations_a}
	\end{subfigure}
	\hfill
	\begin{subfigure}{0.8\linewidth}
		\centering
		\includegraphics[width=\textwidth]{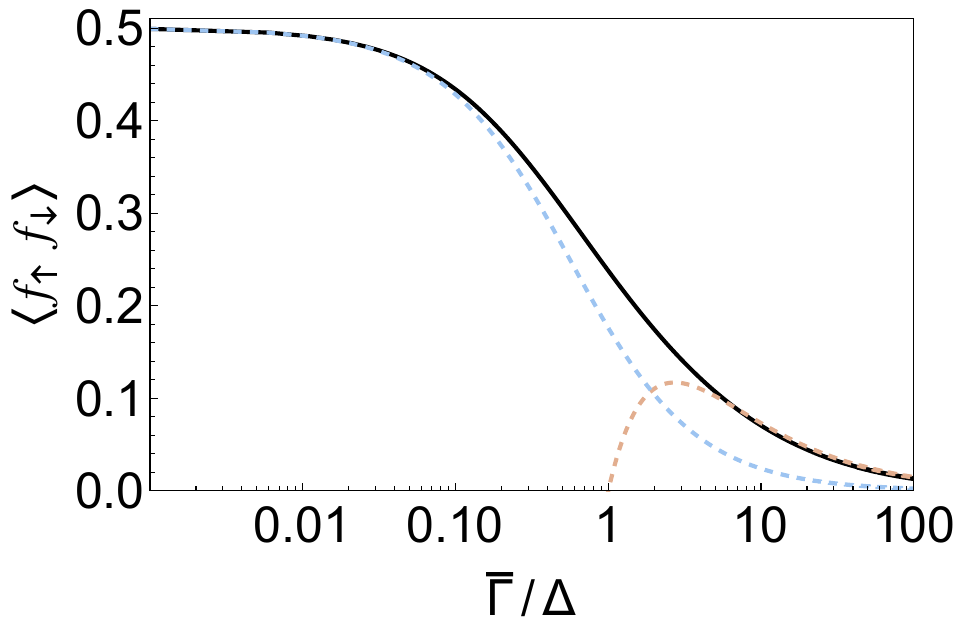}
  		\caption{}
    \label{fig:correlations_b}
	\end{subfigure}
	\caption{\justifying (a) Superconducting correlations as a function of the ratio $\overline{\Gamma}(0)/\Delta,$ with $\overline{\Gamma}(0)$ being the Kondo temperature in the normal state. $\Gamma=0.001W.$ (b) Superconducting correlations as a function of $\overline{\Gamma}/\Delta.$ In this way of plotting the results, the $\pi$ phase gets squished at the zero point of the abscissa. The blue and orange dashed curves are asymptotics for the two opposite limits $\overline{\Gamma}\gg \Delta$ and $\overline{\Gamma}\ll \Delta.$}
     \label{fig:correlations_scaled}
\end{figure}

The quantity $\expval{f_{\uparrow}f_{\downarrow}}$ can be used as a measure of the induced superconducting correlations on the impurity. In the superconducting atomic limit and at zero temperature, the superconducting correlations are equal to the maximal value of $1/2$ in the singlet phase and drop to zero in the magnetic doublet phase.  

The superconducting correlations are given by the anomalous propagator
\begin{equation}
    G^{12}_{ff}(\tau)=-\expval{\mathcal{T}_{\tau}f_{\uparrow}(\tau)f_{\downarrow}(0)},
\end{equation}
with the object of interest being the quantity
\begin{equation}
    \expval{f_{\uparrow}f_{\downarrow}}=-G^{12}_{ff}(0^+).
\end{equation}
Assuming a mean-field decoupling, we have that $G_{ff}(0^+)=B_{xx}(0^+)\mathcal{G}_{dd}(0^+).$ Since $B_{xx}(0^+)=1,$ the superconducting correlations are simply given by $\expval*{f_{\uparrow}f_{\downarrow}}=-\frac{1}{\beta}\sum_{i\omega}\mathcal{G}^{12}_{dd}(i\omega)$ and the auxiliary spin spectral function plays no role in the calculation.

At $T=0^+,$
\begin{equation}\label{eq:correlations}
\expval{f_{\uparrow}f_{\downarrow}}=\int\frac{\dd{\omega}}{2\pi}  \frac{\overline{\Gamma}_\phi\Delta\sqrt{\Delta^2+\omega^2}}{(\overline{\Gamma}_\phi\Delta)^2+\omega^2(\overline{\Gamma}+\sqrt{\Delta^2+\omega^2})^2}.
\end{equation}
The pair correlations are plotted in Fig.~\ref{fig:correlations_a} as a function of the ratio $\overline{\Gamma}(0)/\Delta,$ where $\overline{\Gamma}(0)$ is the renormalized coupling in the absence of superconductivity, given by Eq.~(\ref{eq:TK}). Essentially $\overline{\Gamma}(0)$ plays the role of the Kondo temperature $T_K$ in the normal state, given by its slave-spin mean-field expression. We reproduce the expected result \cite{choi_kondo_2004,luitz_weakcoupling_2010} that in the strong-coupling limit ($T_K\gg \Delta$) an increase in $\Delta$ leads to an increase in the correlations. In the weak-coupling limit ($T_K\ll \Delta$) the pairing correlations are suppressed, instead of exhibiting a jump to negative values. Moreover, an artifact of the slave-spin mean-field is that correlations tend to always saturate at the maximal value $1/2$ before abruptly jumping to zero in the doublet phase.

In Fig.~\ref{fig:correlations_b} we show the superconducting correlations as a function of $\overline{\Gamma}/\Delta.$ With this scaling, all curves collapse upon each other, irrespective of the value of interaction or coupling. This is because we can rescale Eq.~(\ref{eq:correlations}) by $\Delta$, so that, assuming for simplicity that $\phi=0$,
\begin{equation}
\expval{f_{\uparrow}f_{\downarrow}}=\frac{\overline{\Gamma}/\Delta}{2\pi}\int \dd{x}  \frac{\sqrt{x^2+1}}{\pqty{\frac{\overline{\Gamma}}{\Delta}}^2+x^2\pqty{\frac{\overline{\Gamma}}{\Delta}+\sqrt{x^2+1}}^2}.
\end{equation}
If $\overline{\Gamma}\gg \Delta,$ we find that correlations go to zero as $\expval{f_{\uparrow}f_{\downarrow}}\sim \frac{\ln(\overline{\Gamma}/\Delta)}{\pi \overline{\Gamma}/\Delta},$ while if $\overline{\Gamma}\ll \Delta,$ correlations approach the value $1/2$ as $\expval{f_{\uparrow}f_{\downarrow}}\sim \frac{1}{\pi(1+\overline{\Gamma}/\Delta)}\arctan(\frac{1+\overline{\Gamma}/\Delta}{\overline{\Gamma}/\Delta}).$

The $0-\pi$ transition is expected to occur when $T_K/\Delta\sim \mathcal{O}(1),$ while the exact value of proportionality (and definition of $T_K$) differs in the literature, between $T_K/\Delta\sim 0.3$ in the strong coupling limit \cite{Satori_1992,yoshioka_numerical_2000,bauer_spectral_2007} and $T_K/\Delta\sim 0.6$ \cite{lee_scaling_2017}, see also relevant discussion in \cite{gramich_andreev_2017} and citations within. Here, if we equate $T_K=\overline{\Gamma}(0),$ we find the transition at $T_K/\Delta\sim 0.2.$

\section{Microwave response}\label{sec:microwave}

In this section, we use the slave-spin mean-field theory to calculate the linear response of the current to a weak time-dependent modulation of the phase difference across the S-QD-S junction $\phi\to \phi+\delta\phi(t)$, when the QD is in the strongly interacting regime. The motivation is to model such experiments as Fatemi et al.~\cite{fatemi_microwave_2022} where the microwave susceptibility of the junction is measured in a circuit quantum electrodynamic (cQED) setup. To achieve these measurements, the junction is placed in a superconducting ring threaded by a magnetic flux. A small ac modulation of the flux leads to a modulation of the phase difference around the fixed value $\phi,$ resulting in a perturbation of the current.

The main object of interest is the current-current correlation function $\chi_{\mathrm{JJ}}(\omega)$ of the S-QD-S junction in its ground state~\cite{kurilovich_microwave_2021,hermansen_inductive_2022}. By inductively coupling the junction to a microwave resonator, the imaginary part of the current autocorrelation function $\chi''_{\mathrm{JJ}}(\omega)\equiv \Im\chi_{\mathrm{JJ}}(\omega)$ is proportional to the damping rate of the microwave resonator, while the real part $\chi'_{\mathrm{JJ}}(\omega)\equiv \Re\chi_{\mathrm{JJ}}(\omega)$ is proportional to a frequency shift of the resonator. 

There are two contributions to the linear current response. The first is a static part, given by the derivative of the Josephson current, defined in Eq.~(\ref{eq:josephson_current}), with respect to the phase difference, $\pdv{\phi} \expval{J_j}$.  This corresponds to an adiabatic response of the Josephson current. The second contribution is a dynamic part, given by the Kubo formula $\chi(t-t')=-i\theta(t-t')\expval{\comm{J_j(t)}{J_j(t')}}.$ The total response function is
\begin{equation}
    \chi_{\mathrm{JJ}}(\omega)=\pdv{\phi}\expval{J_j}+\delta\chi(\omega),
\end{equation}
where $\delta\chi(\omega)=\chi(\omega)-\chi(0).$ The current autocorrelation function $\chi(\omega)$, in imaginary time, is given by
\begin{equation}
    \chi(\tau)=\expval{J_j(\tau)J_j(0)}-\expval{J_j}^2.
\end{equation}
We take the Josephson current to equal the current from the left lead to the QD ($j=L$). Assuming a mean-field decoupling, we find
\begin{equation}
\begin{split}
    \chi(\tau)\simeq &B_{xx}(\tau)V_j^2 \Tr_N \big[ g_j(\tau)\mathcal{G}_{dd}(-\tau)\\ &+\mathcal{G}_{dd}(\tau)g_j(-\tau) -2 \mathcal{G}_{jd}(\tau)\mathcal{G}_{jd}(-\tau)\big].
\end{split}
\end{equation}
The Fourier transform of this quantity involves the convolution of three propagators, and is calculated to be
\begin{align}
\chi(i\Omega)&=\frac{2V_j^2}{\beta^2}\sum_{i\varepsilon,i\omega}B_{xx}(i\Omega\!+\!i\omega\!-\!i\varepsilon)\times\nonumber\\
&\times\Tr_N \bqty{g_j(i\varepsilon)\mathcal{G}_{dd}(i\omega)-\mathcal{G}_{jd}(i\varepsilon)\mathcal{G}_{jd}(i\omega)}.
\end{align}
For low frequency response ($\Omega\ll 2E$) the sidebands will not contribute significantly (see Appendix~\ref{app:microwave:sidebands} for a justification), so the spin autocorrelation function can be approximated by a central peak around zero frequency, $B_{xx}(i\Omega)\simeq m_x^2 \delta(i\Omega).$ The resulting correlation function is
\begin{equation}\label{eq:chi}
\begin{split}
    \chi(i\Omega)&\simeq \frac{2\overline{V}_j^2}{\beta}\sum_{i\varepsilon}\Tr_N \bqty{g_j(i\Omega\!+\!i\varepsilon)\mathcal{G}_{dd}(i\varepsilon)}\\
    &-\frac{2\overline{V}_j^2}{\beta}\sum_{i\varepsilon}\Tr_N\bqty{\mathcal{G}_{jd}(i\Omega\!+\!i\varepsilon)\mathcal{G}_{jd}(i\varepsilon)}.
\end{split}
\end{equation}
Introducing the relevant spectral functions and performing the analytical continuation $i\Omega\to \Omega+i\eta,$ we obtain the imaginary part of the retarded current-current correlation function
\begin{equation}\label{eq:chiimaginary}
\begin{split}
     \chi''(\Omega)&=-2\pi i\overline{V}_j^2\int\dd{x} \bqty{n_F(x\!-\!\Omega)-n_F(x)}\\ &\times\Tr_N\bqty{\rho_j(x)\rho_d(x\!-\!\Omega)-\rho_{jd}(x)\rho_{jd}(x\!-\!\Omega)}  
\end{split} 
\end{equation}
where matrix elements of spectral functions are defined as $\rho^{ij}(x)=-\Im G^{ij}(x)/\pi$. In what follows, we will distinguish between the two terms on the right-hand side of Eq.~(\ref{eq:chiimaginary}) since the first corresponds to a purely continuum contribution, while the second gives a resonant response due to the ABS states. 

The imaginary part $\chi''(\Omega)$ displays features that correspond to specific absorption processes. We find the same threshold frequencies as in the case of the ground state response of a superconducting weak link~\cite{kos_frequency-dependent_2013,metzger_circuitqed_2021}. The three frequencies are (a) $\Omega=2\Delta,$ corresponding to the creation of two quasiparticles in the superconducting continuum, (b) $\Omega=\Delta+E_{\mathrm{ABS}},$ corresponding to the creation of one quasiparticle in the continuum and another in a bound state, and (c) $\Omega=2E_{\mathrm{ABS}},$ corresponding to the creation of a pair of quasiparticles in one of the ABS. Contrary to the non-interacting weak-link case, however, here $E_{\mathrm{ABS}}$ is the energy of the interacting ABS which can be found through Eq.~(\ref{eq:ABS}). 

\begin{figure}
    \centering
    \includegraphics[width=0.4\textwidth]{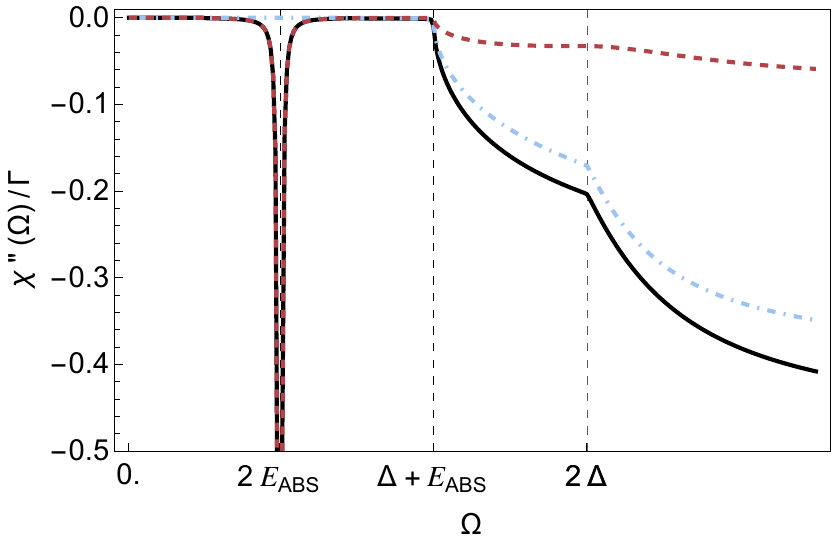}
    \caption{\justifying Imaginary part of the current-current correlation function. The blue dash-dotted line is the contribution from the first term in Eq. (\ref{eq:chiimaginary}), the red dashed line comes from the second term, and the black line is the total response. Parameters are $\Gamma=\Delta=0.01W, U=20\Gamma, \phi=0.$}
    \label{fig:Imchi}
\end{figure}

An illustration of the behavior of the imaginary part is shown in Fig.~\ref{fig:Imchi}, with the three threshold frequencies marked by dashed lines. In Fig.~\ref{fig:Imchi}, we distinguish between two different contributions to the dissipative response: the first term on the right-hand side of Eq.~(\ref{eq:chiimaginary}) (blue dash-dotted line in  Fig.~\ref{fig:Imchi}) corresponds to absorption processes at energies larger than the superconducting gap (cases (a) and (b)), while the second term (red dashed line in  Fig.~\ref{fig:Imchi}) gives a resonant contribution when the frequency of the microwave corresponds to the transition frequency of case (c), as well as a small continuum contribution. Some details and more explicit expressions of the continuum and resonant contributions are given in Appendix~\ref{app:microwave}.

\begin{figure}
    \centering
	\begin{subfigure}{0.8\linewidth}
		\centering
		\includegraphics[width=\textwidth]{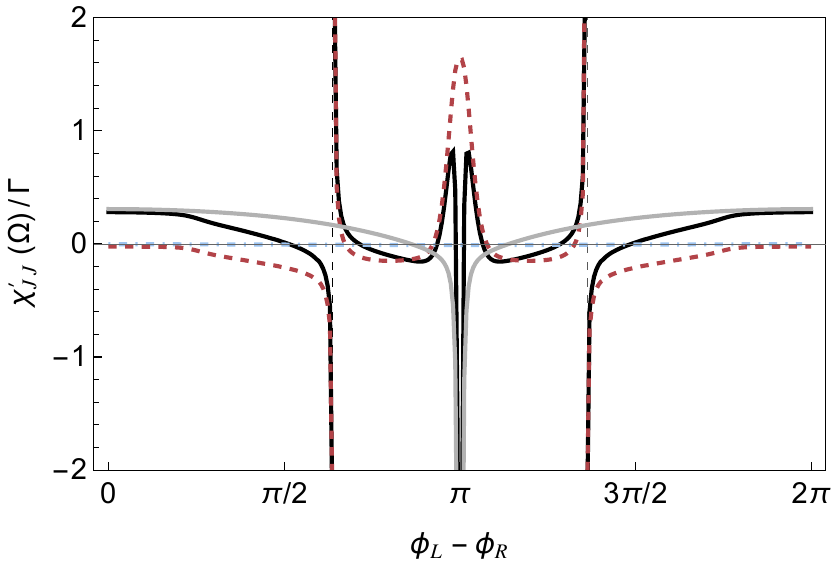}
  		\caption{}
        \label{fig:Rechi:subgap}
	\end{subfigure}
	\begin{subfigure}{0.8\linewidth}
		\centering
		\includegraphics[width=\textwidth]{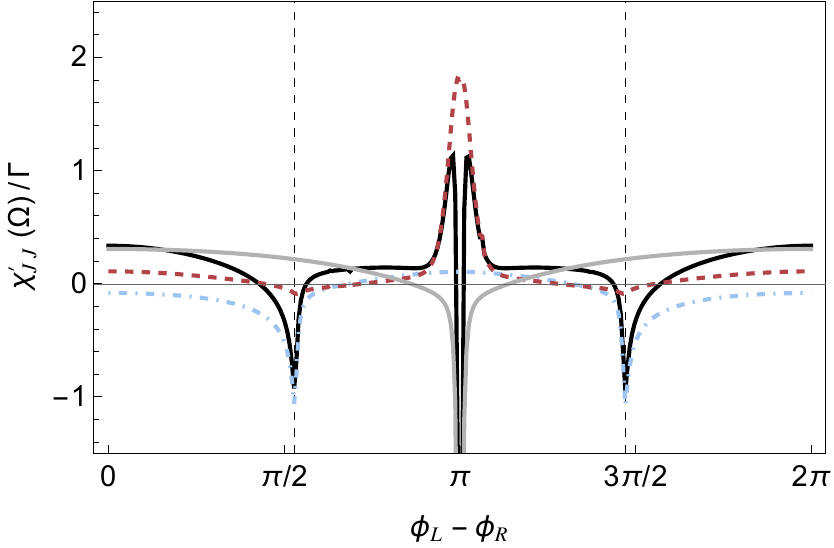}
  		\caption{}
        \label{fig:Rechi:continuum}
	\end{subfigure}
    \caption{\justifying Real part of the total response for resonator frequency (a) $\Omega=0.3\Delta$ and (b) $\Omega=1.2\Delta$. The gray line is the adiabatic response of the Josephson current, while the blue dash-dotted and the red dotted lines correspond to the first and second terms of Eq.~(\ref{eq:chiimaginary}). The vertical dashed lines indicate the values of the phase difference for which (a) $\Omega=2E_{\mathrm{ABS}}(\phi)$ and (b) $\Omega=\Delta+E_{\mathrm{ABS}}(\phi).$ Parameters are $\Gamma=\Delta=0.01W, U=20\Gamma.$}
    \label{fig:Rechi}
\end{figure}
The real part of the total current response $\chi_{JJ}'(\Omega)$ is shown in Fig.~{\ref{fig:Rechi}}. It is convenient to calculate the real part of the susceptibility from the Kramers-Kronig relation
\begin{equation}
\chi'(\Omega)=\pv{\int \frac{\dd{\varepsilon}}{\pi}\frac{\chi''(\varepsilon)}{\varepsilon-\Omega}}
\end{equation}
where $\pv{}$ signifies the Cauchy principal value.

At fixed frequency $\Omega=0.3\Delta$ (Fig.~\ref{fig:Rechi:subgap}), the dominant contribution is the resonant response which produces avoided crossings at values of the phase difference for which $\Omega=2E_{\mathrm{ABS}}(\phi).$ Moreover, one expects that the ABS contribution produces a sharp peak at $\phi_L-\phi_R=\pi$ when the transmission is very large~\cite{metzger_circuitqed_2021,shvetsov_approaching_2025}. We find numerical evidence that the height of this peak is proportional to the renormalized coupling $\overline{\Gamma},$ so that an increasing interaction tends to suppress this peak. The continuum response in Fig.~\ref{fig:Rechi:subgap} is very weak since processes (a) and (b) are not resonant for any value of the phase difference. 

Figure~\ref{fig:Rechi:continuum} shows the response at a frequency larger than the gap, $\Omega=1.2\Delta.$ The response is now dominated by process (b) due to the continuum contribution except at $\phi_L-\phi_R=\pi$ where the resonant response is still large. The continuum contribution produces dips at values of the phase difference for which $\Omega=\Delta+E_{\mathrm{ABS}}(\phi).$

For the regime of parameters chosen, the Josephson current of the QD at zero temperature exhibits a sharp jump at $\phi_L-\phi_R=\pi,$ producing a divergent phase derivative that appears as a large dip in $\chi'_\mathrm{JJ}(\Omega).$ As can be deduced from Fig.~\ref{josephson_current}, the dip is broadened with increasing interaction and moves away from $\phi_L-\phi_R=\pi$ when the junction exhibits a singlet-doublet transition. In experiments, one expects these divergences to be smeared due to finite temperature effects which smoothen the current-phase relation.

Unlike other semi-analytical methods, the slave-spin mean-field method gives access to the microwave response of the superconducting AIM in the Kondo regime. Moreover, it could be a useful tool in regimes where the continuum contribution becomes important, where methods such as the zero-bandwidth approximation fail~\cite{hermansen_inductive_2022}.

\section{Conclusions}\label{sec:conclusions}

In this work, we have discussed the physics of the superconducting Anderson Impurity Model using a slave-spin representation. Within a mean-field decoupling of the spin-fermion Hamiltonian we have obtained the phase diagram and characterized several quantities such as the microwave response, the Josephson current, and the dot spectral function. This analysis reveals that the mean-field theory works qualitatively well in the singlet ($0$) phase, in particular in what concerns the low-energy excitations. However, mean-field misses some important features, most notably giving an incorrect description of the high-energy excitations. Moreover, the mean-field theory predicts a transition into a doublet ($\pi$) phase where the renormalized hybridization vanishes and the system reduces to a decoupled dot. This is particularly evident in the spectral function, where the Hubbard bands are renormalized to zero. At the mean-field level, all the physical properties of the doublet phase are missed.

To go beyond this approach, we have included RPA corrections to the mean-field treatment of the spin sector, leading to a finite dissipation of the spin dynamics and hence a finite width of the Hubbard bands. This also has direct consequences on other observables. Note that within our approach, the RPA corrections to the mean-field theory do not renormalize the $0-\pi$ transition boundary, nor are they able to give a description of the low-energy physics in the $\pi$ phase. In order to access the $\pi$ phase, and in particular the ABS, one would likely need to go beyond the fermion-spin decoupling at the Green's function level, namely introduce vertex corrections in the calculation of the Green's functions.

As a non-trivial application of the method, we compute the microwave spectroscopy of the model in the strongly interacting regime, which is a quantity of direct experimental relevance. We show that it displays a rich structure, with resonant and continuum contributions.

Some interesting extensions to the present method would be to do a self-consistent approximation, where the position of the critical boundary is also renormalized, or to extend the method to the doublet phase. Moreover, it should be straightforward to extend the method outside particle-hole symmetry, following~\cite{guerci_unbinding_2017}. The present method could be a promising way to semi-analytically study more complicated setups such as multiterminal QD junctions in the presence of interactions, such as the minimal Kitaev chain or the Andreev molecule~\cite{pillet_josephson_2023,kocsis_strong_2024}. Such multi-impurity problems have rich phase diagrams~\cite{zalom_doublequantumdot_2024} which would be interesting to explore. Future perspectives include the description of nonequilibrium scenarios, including time-periodic driving~\cite{keliri2023driven,keliri2023longrange} and transport dynamics.

\begin{acknowledgments}
We would like to thank L. Bretheau and J. D. Pillet for discussions on this work. We acknowledge financial support from the ERC consolidator grant No.~101002955 - CONQUER.
\end{acknowledgments}

\appendix

\section{Slave-spin approach for $\Delta\rightarrow\infty$}
\label{sec:largegap}

It is useful to briefly recap the large gap limit $\Delta\rightarrow\infty$ and how one can describe it within the slave-spin framework. In this limit, one can derive an effective model for the dot~\cite{meng_selfconsistent_2009}, featuring Coulomb repulsion and local pairing
\begin{align}
H=-\vert\Gamma_{\varphi}\vert\left(f^{\dagger}_{\uparrow}f^{\dagger}_{\downarrow}+\mathrm{H.c.}\right)+\frac{U}{2}(n-1)^2.
\end{align}
In the slave-spin representation of the model, we see that the local pairing is not affected by the auxiliary two-level variable. We have indeed
\begin{align}
H'=-\vert\Gamma_{\varphi}\vert\left(d^{\dagger}_{\uparrow}d^{\dagger}_{\downarrow}+\mathrm{H.c.}\right)+\frac{U}{4}(1-\sigma^z).
\end{align}
The spectrum and eigenstates of the model in the slave-spin formulation are easily obtained. In the physical subspace satisfying the constraint $\sigma^z\Omega=1$ the states $\ket{\uparrow}\ket{+},\ket{\downarrow}\ket{+}$ are degenerate with energy $E=0$, while the Bogoliubov states $\ket{\phi_{\pm}}\ket{-}$
have energy $E_{\pm}=U/2\pm \vert\Gamma_{\varphi}\vert $, where at half-filling we have $\ket{\phi_{\pm}}=(\ket{0} \pm \ket{\uparrow\downarrow})/2 $. As the pairing strength increases at fixed $U$, the system undergoes a level crossing from a doublet state $\ket{\uparrow}\ket{+},\ket{\downarrow}\ket{+}$ to the BCS singlet state $\ket{\phi_{-}}\ket{-}$. At the level crossing, the fraction of doublons jumps from $D=\frac{1-\sigma^z}{4}=0$ to $D=1/2$.
Most importantly, we see that even in the doublet phase the effective superconducting pairing remains finite, as opposed to the case of mean-field decoupling in which it vanishes at the transition, as discussed in the main text.

\section{Abrikosov pseudofermion Green's function in mean-field}\label{app:mf}

Introducing Abrikosov pseudofermions, we rewrite the slave-spin Hamiltonian (\ref{Htotal}) as

\begin{equation}
\begin{split}
  \mathcal{H}'&=\mathcal{H}_{\mathrm{BCS}}-u\sum_{\alpha\beta}\psi^{\dagger}_\alpha \sigma^z_{\alpha\beta}\psi_\beta \\
  &+ \sum_{jk\sigma}\sum_{\alpha\beta}V \pqty{c^{\dagger}_{jk\sigma}\psi^{\dagger}_\alpha \sigma^x_{\alpha\beta}\psi_\beta d_\sigma +\mathrm{H.c.}}
\end{split}  
\end{equation}
where $u\equiv U/4.$ If the pseudofermions are free ($V=0$), we obtain the bare propagator of the $\psi$ fermions,
\begin{equation}\label{baregreen}
G^0_\psi(i\omega)=(i\omega+u\sigma^z)^{-1},
\end{equation}
with $i\omega$ being fermionic Matsubara frequencies. Note that when working at finite temperature we follow the Popov-Fedotov approach which adds a chemical potential to the Abrikosov pseudofermions $\mu^f=i\frac{\pi}{2\beta}$ \cite{popov_functionalintegration_1988,kiselev_schwingerkeldysh_2000,brinckmann_diagrammatic_2008}.

For the interacting problem ($V\neq0$), the equation of motion for the $\psi$ fermions is
\begin{equation}
\begin{split}
    -\dv{\tau}G^{\alpha\beta}_\psi(\tau)&=\delta(\tau) \expval{\acomm{\psi_\alpha}{\psi^\dagger_\beta}}+\expval{T_\tau \comm{\mathcal{H}(\tau)}{\psi_\alpha(\tau)}\psi^\dagger_\beta}\\
    &=\delta(\tau)\delta_{\alpha\beta}-u\sum_{\beta'}\sigma^z_{\alpha\beta'}G^{\beta'\beta}_\psi\\
    +&\sum_{jk\sigma}\sum_{\beta'}V\sigma^x_{\alpha\beta'}
\expval{T_\tau (c^{\dagger}_{jk\sigma}\psi_{\beta'} d_\sigma+\mathrm{H.c.})(\tau) \psi^\dagger_\beta}.
    \end{split}
 \end{equation}
Using a Hartree-Fock decoupling, we approximate the last term by $\expval{T_\tau (c^{\dagger}\psi_{\beta'} d)(\tau) \psi^\dagger_\beta}\simeq -\expval{c^\dagger d}_{F}G_\psi^{\beta'\beta}.$ This defines a self-energy $\Sigma_\psi=\sigma^x h,$ where $h$ is the average hybridization defined in Eq. (\ref{hdefinition}). We see that
\begin{equation}\label{gpsi}
\begin{split}
   G_\psi(i\omega)& =\frac{1}{i\omega+u\sigma^z-h\sigma^x}=\frac{i\omega-u\sigma^z+h\sigma^x}{(i\omega)^2-E^2},
    \end{split}
\end{equation}
with $E^2=h^2+u^2.$ Equivalently, we can write the Dyson equation $G_\psi=G_\psi^0+G_\psi^0\Sigma_\psi G_\psi.$

\subsection{Expectation values of spin operators}
The average values of spin operators can be easily calculated using the relation
\begin{equation}
   \expval{\sigma^i}=\sum_{\alpha\beta}\expval{\psi^\dagger_\alpha \sigma^i_{\alpha\beta}\psi_\beta}=\Tr_{\sigma}(\sigma^i G_\psi(0^-)),
\end{equation}
where the trace is over the spin degrees of freedom. Inserting Eq. (\ref{gpsi}) in the above expression we get, e.g. for the magnetization along $x$:
\begin{equation}
\begin{split}
    \expval{\sigma^x}&=\frac{1}{\beta}\sum_{i\omega} e^{i\omega 0^+}\Tr_{\sigma}(\sigma^x G_\psi(i\omega))\\
    &=\frac{1}{\beta}\sum_{i\omega} e^{i\omega 0^+}\frac{2h}{(i\omega+\mu^f-E)(i\omega+\mu^f+E)}\\
    &=\frac{h}{E}\bqty{n_F(E-\mu^f)-n_F(-E-\mu^f)}\\
    &=-\frac{h}{E}\tanh(\beta E)
    \end{split}
\end{equation}
and, \textit{mutatis mutandis} for the magnetization along $z,$
\begin{equation}
  \expval{\sigma^z}=\frac{u}{E}\tanh(\beta E). 
\end{equation}
In the above calculations, a useful relation for the Fermi function is
\begin{equation}\label{useful}
    n_F(x-\mu^f)=n_F(2x)+i\frac{1}{2\cosh(\beta x)}.
\end{equation}

\subsection{Spin-spin correlation function in mean-field}\label{app:bxx}
The spin-spin correlation function is defined as
\begin{equation}
    B_{ij}(\tau)=\sum_{\alpha\beta\gamma\delta}\expval{T_\tau \psi^{\dagger}_\alpha(\tau) \sigma^i_{\alpha\beta}\psi_\beta(\tau)\psi^{\dagger}_\gamma\sigma^j_{\gamma\delta}\psi_\delta}.
\end{equation}
In the non-interacting case we can use Wick's theorem to write
\begin{equation}
  B^0_{ij}(\tau)=\expval{\sigma^i}\expval{\sigma^j}-\Tr_{\sigma}\bqty{\sigma^i G^0_\psi(\tau)\sigma^j G^0_\psi(-\tau)}.
\end{equation}
The Fourier expansion in Matsubara frequencies turns the product of the two Green's functions into a convolution
\begin{equation}
    \int_0^\beta \dd{\tau} e^{i\Omega \tau}G(\tau)G(-\tau)=\frac{1}{\beta}\sum_{\nu}G(i\Omega+i\nu)G(i\nu),
\end{equation}
resulting in
\begin{equation}
\begin{split}
    B^0_{ij}(i\Omega)&=\expval{\sigma^i}\expval{\sigma^j}\delta(i\Omega)\\
    &-\frac{1}{\beta}\sum_{i\nu}\Tr_{\sigma}\bqty{\sigma^i G^0_\psi(i\Omega+i\nu)\sigma^j G^0_\psi(i\nu)},
\end{split}
\end{equation}
with $\Omega$ representing bosonic and $\nu$ representing fermionic Matsubara frequencies. We define the incoherent contribution as
\begin{equation}
\Pi^0_{ij}(i\Omega)=-\frac{1}{\beta}\sum_{i\nu}\Tr_\sigma\!\!\bqty{\sigma^iG^0_\psi(i\Omega+i\nu)\sigma^j G^0_\psi(i\nu)}.
\end{equation}
In order to obtain the spin autocorrelation function in mean-field we dress the pair-bubble by replacing the bare propagator $G^0_\psi$ with $G_\psi,$ which is given by Eq. (\ref{gpsi}). We find
\begin{equation}
\begin{split}
    \Pi_{xx}(i\Omega)&=-\frac{1}{\beta}\sum_{i\nu}\Tr_{\sigma}\!\!\bqty{\sigma^x G_\psi(i\Omega+i\nu)\sigma^x G_\psi(i\nu)} \\
    &=(1-m_x^2)\tanh(\beta E)\\ &\times\pqty{\frac{1}{i\Omega+2E}-\frac{1}{i\Omega-2E}}.
\end{split}
\end{equation}
This concludes the calculation of the mean-field spin-spin correlation function,
\begin{equation}
B_{xx}(i\Omega)=m_x^2\delta(i\Omega)+\Pi_{xx}(i\Omega).
\end{equation}

\section{Analytical continuation of the self-energy}\label{app:rpa}
The goal of this section is to show how to analytically continue the spin self-energy $\Sigma_{xx}$ from Matsubara to real frequencies, as well as to justify the approximations used in the calculation of the spectral function in Fig.~\ref{fig:specRPA_a}. We start from a rewriting of Eq.~(\ref{eq:sigmamatsubara}),
\begin{equation}
  \begin{split}
    \Sigma_{xx}(i\Omega)&=\frac{-2V^2}{\beta}\Tr_N\sum_{i\nu}\sum_j g_j(i\nu)\mathcal{G}_{dd}(i\Omega+i\nu)\\
    &-\frac{2V^2}{\beta}\Tr_N\sum_{i\nu}\sum_{jj'}\mathcal{G}_{jd}(i\nu)\mathcal{G}_{j'd}(i\Omega+i\nu).
\end{split}  
\end{equation}
The second term  is of the order of $\Gamma^2 m_x^2$ so that at large $U/\Gamma$ values is much smaller than the first term, which is of order $\Gamma.$ We therefore ignore the contribution from the second term in what follows. Introducing the spectral representations of the relevant Green's functions, the self-energy is written as

\begin{equation}
\begin{split}
    \Sigma_{xx}(i\Omega)&\simeq\frac{-2V^2}{\beta}\Tr_N\sum_{i\nu} \sum_j g_j(i\nu)\mathcal{G}_{dd}(i\Omega+i\nu)\\
    &=-2V^2\sum_j\iint\dd{x}\dd{y} \frac{\Tr_N\bqty{\rho_j(x)\rho_d(y)}}{i\Omega+x-y} \\ &\times\bqty{n_F(x)-n_F(y)}.
\end{split}
\end{equation}

We perform the analytical continuation $i\Omega\to \Omega+i\eta,$ and find the real and imaginary parts of the above quantity:

\begin{equation}
\begin{split}
   \Re\Sigma_{xx}(\Omega)&=-2V^2\iint\dd{x}\dd{y}\Tr_N\sum_j\rho_j(x)\rho_d(y)\\
   &\times \frac{\principalvalue}{\Omega+x-y}\bqty{n_F(x)-n_F(y)},
   \end{split}
\end{equation}

and

\begin{equation}
\begin{split}
\Im\Sigma_{xx}(\Omega)&=2\pi V^2\int\dd{x}\bqty{n_F(x)-n_F(x+\Omega)}\\\ &\times \Tr_N\sum_j\rho_j(x)\rho_d(x+\Omega).
\end{split}
\end{equation}

At zero temperature an explicit expression for the imaginary part is
\begin{equation}
\begin{split}
\Im\Sigma_{xx}(\Omega)&=4\Gamma\mathrm{sign}(\Omega)\int_\Delta^{\abs{\Omega}}\dd{x}\frac{\theta(\abs{\Omega}-\Delta)}{\sqrt{x^2-\Delta^2}}\\ &\times \bqty{x\rho_d^{11}(\abs{\Omega}\!-\!x)\!-\!\Delta\cos\phi\rho_d^{12}(\abs{\Omega}\!-\!x)},
\end{split}
\end{equation}
where $\theta(x)$ is the Heaviside step function. This term has an extended gap, $\abs{\Omega}>\Delta+E_{\mathrm{ABS}},$ and, in the limit $\Omega\gg\Delta,$ reduces to $\Im\Sigma_{xx}(\Omega)=2\Gamma \mathrm{sign}(\Omega).$ We find that the real part at large frequencies behaves as $-(4\Gamma/\pi)\ln\abs{\Omega/W}$ and is therefore generally much smaller compared to the Hubbard band energy scale which is given by the interaction $U.$ In the limit $U\gg \Gamma$ we can safely ignore any shift of the Hubbard bands produced by $\Re\Sigma_{xx}.$

\section{Details on the microwave response}\label{app:microwave}
Starting from the imaginary part of the current-current correlation function at zero temperature, we decompose it into its two contributions for positive frequencies $\Omega>0$
\begin{align}
     &\chi_1''(\Omega)=-2\pi i\overline{V}_j^2\int_0^\Omega\dd{x} \Tr_N\bqty{\rho_j(x)\rho_d(x\!-\!\Omega)}\\
    &\chi_2''(\Omega)=2\pi i\overline{V}_j^2\int_0^\Omega\dd{x} \Tr_N\bqty{\rho_{jd}(x)\rho_{jd}(x\!-\!\Omega)}
\end{align}
where the first is a purely continuum contribution and the second gives mainly a resonant contribution. We provide some details on these two terms in the following subsections.
\subsection{Continuum contribution}
A more explicit expression for the first term is
\begin{equation}
\begin{split}
    \chi''_1(\Omega)&=-2\overline{\Gamma}\int_{\Delta}^{\Omega}\dd{x} \frac{\theta(\Omega\!-\!\Delta)}{\sqrt{x^2\!-\!\Delta^2}}\\
    &\times \bqty{x\rho_d^{11}(x\!-\!\Omega)+\Delta\cos\phi \rho_d^{12}(x\!-\!\Omega)}.  
\end{split}
\end{equation}
This term has an extended gap and contributes only if $\Omega\geq\Delta+E_{\mathrm{ABS}}.$ A second absorption process produces a cusp at frequencies $\Omega=2\Delta.$ At large frequencies $\Omega\gg\Delta,$ $\chi''_1(\Omega)$ decays to $-\overline{\Gamma}.$ The corresponding real part of the susceptibility $\delta\chi_1'(\Omega)$ has resonant features at the values of the phase difference $\phi$ for which $\Omega=\Delta+E_{\mathrm{ABS}}(\phi).$ 

For the calculation of $\delta\chi'_1(\omega)$ a cutoff frequency $\omega_c$ is introduced. Then, the correction error is $-\frac{\overline{\Gamma}}{\pi}\log\frac{\omega_c^2}{\omega_c^2-\omega^2}$ and 
\begin{equation}
\begin{split}
      \delta\chi_1'(\Omega)&=\int_\Delta^{\omega_c}\frac{\dd{\varepsilon}}{\pi}\chi_1''(\varepsilon)\bqty{\frac{2\varepsilon}{(\varepsilon\!-\!\Omega)^2+\eta^2}-\frac{2\varepsilon}{\varepsilon^2+\eta^2}}\\ &-\frac{\overline{\Gamma}}{\pi}\log\frac{\omega_c^2}{\omega_c^2-\omega^2}.
\end{split}
\end{equation}

\subsection{Resonant contribution}
The imaginary part of the $\chi_2(\Omega)$ term has a resonant contribution at $\Omega=2E_{\mathrm{ABS}}$ and a small continuum contribution for frequencies above the gap. Keeping only the resonant term is equivalent to considering the case $\Omega\ll\Delta,$ and we can then approximate the resonant contribution by
\begin{equation}
\begin{split}
   \chi''_{r}(\Omega)=-\pi \overline{\Gamma}^2 \int_0^\Omega\dd{x} \big[\rho^{11}_d(x)\rho^{11}_{d}(x\!-\!\Omega)&\\
   -\cos 2\phi\, \rho^{12}_d(x)\rho^{12}_{d}(x\!-\!\Omega)\big]&. 
\end{split}  
\end{equation}
We can fit $ \chi''_{r}(\Omega)$ with a Lorentzian of width $\gamma,$ provided that $E_\mathrm{ABS}$ is not close to zero (the fit is therefore not good close to $\phi_L-\phi_R=\pi$)
\begin{equation}
\chi''_r(\Omega)\simeq \pi \gamma \chi''_r(2E_{\mathrm{ABS}})\delta(\Omega-2E_{\mathrm{ABS}}).
\end{equation}
We find that the corresponding real part of the response 
\begin{equation}
\begin{split}
\chi'_{r}(\Omega)&=\gamma \chi''_{r}(2E_{\mathrm{ABS}})\\
       &\times\Re\bqty{\frac{1}{2E_{\mathrm{ABS}}\!-\!\Omega\!-\!i\eta}+\frac{1}{2E_{\mathrm{ABS}}\!+\!\Omega\!+\!i\eta}}
\end{split}
\end{equation}
changes sign at values of the phase for which $\Omega=2E_\mathrm{ABS},$ and therefore exhibits an avoided crossing.

\subsection{High-frequency contribution}\label{app:microwave:sidebands}
The contribution from the Hubbard bands is
\begin{equation}
\begin{split}
  \chi_{\mathrm{inc}}(i\Omega)&=\frac{2V_j^2}{\beta^2}\sum_{i\varepsilon,i\omega}\Pi_{xx}(i\Omega\!+\!i\varepsilon\!-\!i\omega)\\ &\times\Tr_N \bqty{g_j(i\omega)\mathcal{G}_{dd}(i\varepsilon)-\mathcal{G}_{jd}(i\omega)\mathcal{G}_{jd}(i\varepsilon)}.   
\end{split}
\end{equation}
We calculate the sideband contributions at $T=0$ and positive frequencies $\Omega>0,$ and get
\begin{equation}
\begin{split}
    \chi''_{\mathrm{inc}}(\Omega)&=\frac{\Gamma}{\rho_0}\int_\Delta^\Omega\dd{x}\int_0^{\Omega}\dd{y}\Tr_N\bqty{\rho_j(x)\rho_d(-y)}\\ &\times \theta(\Omega\!-\!x\!-\!y)\rho_s(\Omega\!-\!x\!-\!y)\\
    &-\frac{\Gamma}{\rho_0}\int_0^\Omega\dd{x}\int_0^{\Omega}\dd{y}\Tr_N\bqty{\rho_{jd}(x)\rho_{jd}(-y)}\\ &\times  \theta(\Omega\!-\!x\!-\!y) \rho_s(\Omega\!-\!x\!-\!y)
\end{split}  
\end{equation}  
Both terms on the right-hand side only contribute at high frequencies; the first term for $\Omega\gtrsim  \Delta+E_{\mathrm{ABS}}+2E$ and the second for $\Omega\gtrsim 2E_{\mathrm{ABS}}+2E.$

The first term is similar to a smeared step function. It saturates to $\Gamma(1-m_x^2)$ at large frequencies $\Omega\gg 2E.$ The corresponding real part can be approximated by
\begin{equation}
   \chi'_{1,\mathrm{inc}}(\Omega)\simeq \frac{\Gamma(1\!-\!m_x^2)}{\pi}\log\frac{(2E+\Delta+E_{\mathrm{ABS}})^2}{(2E\!+\!\Delta\!+\!E_{\mathrm{ABS}})^2\!-\!\Omega^2}
\end{equation}
For low frequency response $\Omega\ll E$, the contribution due to this term is negligible $\chi'_{1,\mathrm{inc}}(\Omega\ll E)\to 0.$ The second term behaves like $\chi''_{2,\mathrm{inc}}(\Omega)\propto\Gamma(1-m_x^2)\theta(\Omega-2E-2E_{\mathrm{ABS}})/2\pi\Omega $ and also has a negligible contribution to the low frequency response.

\bibliography{references,bib}

\end{document}